\documentclass[twocolumn,aps,prd,nofootinbib,superscriptaddress,showpacs]{revtex4-1}
\usepackage{amsmath,graphicx,bm,amsbsy,aas_macros,color,natbib,subfigure,enumerate}
\pdfoutput=1

\newcommand{\modelResidualCorr}{0.116}
\newcommand{\limitDeltaSq}{3.7\times 10^4}
\newcommand{\limitk}{0.18}
\newcommand{\limitz}{6.8}
\newcommand{\wedgeBuffer}{0.02}

\newcommand{\M}{\mathbf{M}}
\newcommand{\C}{\mathbf{C}}
\newcommand{\beq}{\begin{equation}}
\newcommand{\eeq}{\end{equation}}

\newcommand{\x}{\mathbf{x}}
\newcommand{\kvec}{\mathbf{k}}

\newcommand{\mean}{\boldsymbol\mu}

\newcommand{\p}{\mathbf{p}}

\newcommand{\Proj}{\mathbf{\Pi}}
\newcommand{\PSF}{\mathbf{P}}

\newcommand{\trans}{\mathsf{T}}

\newcommand{\xhat}{\widehat{\x}}
\newcommand{\CHat}{\widehat{\C}}

\def\ASU{School of Earth and Space Exploration, Arizona State University, Tempe, AZ 85287, USA}
\def\ANU{Research School of Astronomy and Astrophysics, Australian National University, Canberra, ACT 2611, Australia}
\def\CASS{CSIRO Astronomy and Space Science (CASS), PO Box 76, Epping, NSW 1710, Australia}
\def\Curtin{International Centre for Radio Astronomy Research, Curtin University, Bentley, WA 6102, Australia}
\def\CfA{Harvard-Smithsonian Center for Astrophysics, Cambridge, MA 02138, USA}
\def\Haystack{MIT Haystack Observatory, Westford, MA 01886, USA}
\def\MITKavli{MIT Kavli Institute, Massachusetts Institute of Technology, Cambridge, MA, 02139 USA}
\def\MITPhysics{Department of Physics, Massachusetts Institute of Technology, Cambridge, MA, 02139 USA}
\def\RRI{Raman Research Institute, Bangalore 560080, India}
\def\SKASA{Square Kilometre Array South Africa (SKA SA), Cape Town 7405, South Africa}
\def\Tata{National Centre for Radio Astrophysics, Tata Institute for Fundamental Research, Pune 411007, India}
\def\UMelbourne{School of Physics, The University of Melbourne, Parkville, VIC 3010, Australia}
\def\USydney{Sydney Institute for Astronomy, School of Physics, The University of Sydney, NSW 2006, Australia}
\def\UW{Department of Physics, University of Washington, Seattle, WA 98195, USA}
\def\UWisc{Department of Physics, University of Wisconsin--Milwaukee, Milwaukee, WI 53201, USA}
\def\Victoria{School of Chemical \& Physical Sciences, Victoria University of Wellington, Wellington 6140, New Zealand}
\def\CAASTRO{ARC Centre of Excellence for All-sky Astrophysics (CAASTRO)}
\def\Rhodes{Department of Physics and Electronics, Rhodes University, Grahamstown, 6140, South Africa}

\begin{document} 

\title{Empirical Covariance Modeling for 21\,cm Power Spectrum Estimation: A Method Demonstration and New Limits from Early Murchison Widefield Array 128-Tile Data}

\author{Joshua~S.~Dillon}
\email{jsdillon@mit.edu}
\affiliation{\MITPhysics}\affiliation{\MITKavli}

\author{Abraham~R.~Neben}
\affiliation{\MITPhysics}\affiliation{\MITKavli}

\author{Jacqueline~N.~Hewitt}
\affiliation{\MITPhysics}\affiliation{\MITKavli}

\author{Max Tegmark}
\affiliation{\MITPhysics}\affiliation{\MITKavli}

\author{N.~Barry}
\affiliation{\UW}

\author{A.~P.~Beardsley}
\affiliation{\UW}

\author{J.~D.~Bowman}
\affiliation{\ASU}

\author{F.~Briggs}
\affiliation{\ANU} \affiliation{\CAASTRO}

\author{P.~Carroll}
\affiliation{\UW}

\author{A.~de~Oliveira-Costa}

\author{A.~Ewall-Wice}
\affiliation{\MITPhysics}\affiliation{\MITKavli}

\author{L.~Feng}
\affiliation{\MITPhysics}\affiliation{\MITKavli}


\author{L.~J.~Greenhill}
\affiliation{\CfA}

\author{B.~J.~Hazelton}
\affiliation{\UW}

\author{L.~Hernquist}
\affiliation{\CfA}

\author{N.~Hurley-Walker}
\affiliation{\Curtin}

\author{D.~C.~Jacobs}
\affiliation{\ASU}

\author{H.~S.~Kim}
\affiliation{\UMelbourne}\affiliation{\CAASTRO}

\author{P.~Kittiwisit}
\affiliation{\ASU}

\author{E.~Lenc}
\affiliation{\USydney}\affiliation{\CAASTRO}

\author{J.~Line}
\affiliation{\UMelbourne}\affiliation{\CAASTRO}

\author{A.~Loeb}
\affiliation{\CfA}

\author{B.~McKinley}
\affiliation{\UMelbourne}\affiliation{\CAASTRO}

\author{D.~A.~Mitchell}
\affiliation{\CASS} \affiliation{\CAASTRO}

\author{M.~F.~Morales}
\affiliation{\UW}

\author{A.~R.~Offringa}
\affiliation{\ANU} \affiliation{\CAASTRO}

\author{S.~Paul}
\affiliation{\RRI}

\author{B.~Pindor}
\affiliation{\UMelbourne}\affiliation{\CAASTRO}

\author{J.~C.~Pober}
\affiliation{\UW}

\author{P.~Procopio}
\affiliation{\UMelbourne}\affiliation{\CAASTRO}

\author{J.~Riding}
\affiliation{\UMelbourne}\affiliation{\CAASTRO}

\author{S.~Sethi}
\affiliation{\RRI}

\author{N.~Udaya~Shankar}
\affiliation{\RRI}

\author{R.~Subrahmanyan}
\affiliation{\RRI} \affiliation{\CAASTRO}

\author{I.~Sullivan}
\affiliation{\UW}

\author{Nithyanandan~Thyagarajan}
\affiliation{\ASU}

\author{S.~J.~Tingay}
\affiliation{\Curtin} \affiliation{\CAASTRO}

\author{C.~Trott}
\affiliation{\Curtin} \affiliation{\CAASTRO}

\author{R.~B.~Wayth}
\affiliation{\Curtin} \affiliation{\CAASTRO}

\author{R.~L.~Webster}
\affiliation{\UMelbourne} \affiliation{\CAASTRO}

\author{S.~Wyithe}
\affiliation{\UMelbourne} \affiliation{\CAASTRO}

\author{G.~Bernardi}
\affiliation{\Rhodes}\affiliation{\SKASA} 

\author{R.~J.~Cappallo}
\affiliation{\Haystack}
 
\author{A.~A.~Deshpande}
\affiliation{\RRI}

\author{M.~Johnston-Hollitt}
\affiliation{\Victoria}

\author{D.~L.~Kaplan}
\affiliation{\UWisc}

\author{C.~J.~Lonsdale}
\affiliation{\Haystack}

\author{S.~R.~McWhirter}
\affiliation{\Haystack}

\author{E.~Morgan}
\affiliation{\MITKavli}

\author{D.~Oberoi}
\affiliation{\Tata}

\author{S.~M.~Ord}
\affiliation{\Curtin} \affiliation{\CAASTRO}

\author{T.~Prabu}
\affiliation{\RRI}

\author{K.~S.~Srivani}
\affiliation{\RRI}

\author{A.~Williams}
\affiliation{\Curtin}

\author{C.~L.~Williams}
\affiliation{\MITPhysics}\affiliation{\MITKavli}

\date{May 22, 2015; Submitted: March 10, 2015}



\begin{abstract}
The separation of the faint cosmological background signal from bright astrophysical foregrounds remains one of the most daunting challenges of mapping the high-redshift intergalactic medium with the redshifted 21\,cm line of neutral hydrogen. Advances in mapping and modeling of diffuse and point source foregrounds have improved subtraction accuracy, but no subtraction scheme is perfect.  Precisely quantifying the errors and error correlations due to missubtracted foregrounds allows for both the rigorous analysis of the 21\,cm power spectrum and for the maximal isolation of the ``EoR window'' from foreground contamination. We present a method to infer the covariance of foreground residuals from the data itself in contrast to previous attempts at \textit{a priori} modeling. We demonstrate our method by setting limits on the power spectrum using a 3\,h integration from the 128-tile Murchison Widefield Array. Observing between 167 and 198\,MHz, we find at 95\% confidence a best limit of $\Delta^2(k) < \limitDeltaSq$\,mK$^2$ at comoving scale $k = \limitk$\,$h$\,Mpc$^{-1}$ and at $z = \limitz$, consistent with existing limits.
\end{abstract}

\maketitle
 

\section{Introduction} \label{sec:intro}

Tomographic mapping of neutral hydrogen using its 21\,cm hyperfine transition has the potential to directly probe the density, temperature, and ionization of the intergalactic medium (IGM), from redshift 50 (and possibly earlier) through the end of reionization at $z\sim 6$. This unprecedented view of the so-called ``cosmic dawn'' can tightly constrain models of the first stars and galaxies \cite{FurlanettoReview, miguelreview, PritchardLoebReview, aviBook} and eventually yield an order of magnitude more precise test of the standard cosmological model ($\Lambda$CDM) than current probes \cite{Yi}. 

Over the past few years, first generation instruments have made considerable progress toward the detection of the power spectrum of the 21\,cm emission during the epoch of reionization (EoR). Telescopes such as the Low Frequency Array (LOFAR \cite{LOFARinstrument}), the Donald C. Backer Precision Array for Probing the Epoch of Reionization (PAPER \cite{PAPER}), the Giant Metrewave Radio Telescope (GMRT \cite{newGMRT}), and the Murchison Widefield Array (MWA \cite{MWAdesign,TingaySummary,BowmanMWAScience}) are now operating, and have begun to set limits on the power spectrum. GMRT set some of the earliest limits \cite{newGMRT} and both PAPER \cite{DannyMultiRedshift} and the MWA \cite{X13} have presented upper limits across multiple redshifts using small prototype arrays. PAPER has translated its results into a constraint on the heating of the IGM by the first generation of x-ray binaries and miniquasars \cite{PAPER32Limits} and has placed the tightest constraints so far on the power spectrum \cite{PAPER64Limits} and the thermal history of the IGM \cite{PoberPAPER64Heating}.

Despite recent advances, considerable analysis challenges remain. Extracting the subtle cosmological signal from the noise is expected to require thousand hour observations across a range of redshifts \citep{MiguelNoise,Judd06,LidzRiseFall,LOFAR2,AaronSensitivity,ThyagarajanWedge}. Even more daunting is the fact that the 21\,cm signal is probably at least 4 orders of magnitude dimmer than the astrophysical foregrounds---due to synchrotron radiation from both our Galaxy and from other galaxies \citep{Angelica,LOFAR,BernardiForegrounds,PoberWedge,InitialLOFAR1,InitialLOFAR2}. 

Recently, simulations and analytical calculations have established the existence of a region in cylindrical Fourier space---in which three-dimensional (3D) Fourier modes $\vec{k}$ are binned into $k_\|$ modes along the line of sight and  $k_\perp$ modes perpendicular to it---called the ``EoR window'' that should be fairly free of foreground contamination \cite{Dattapowerspec,AaronDelay,VedanthamWedge,MoralesPSShapes,Hazelton2013,CathWedge,ThyagarajanWedge,EoRWindow1,EoRWindow2}. Observations of the EoR window confirm that it is largely foreground-free \cite{PoberWedge,X13} up to the sensitivity limits of current experiments. The boundary of the EoR window is determined by the volume and resolution of the observation, the intrinsic spectral structure of the foregrounds, and the so-called ``wedge.'' 

Physically, the wedge arises from the frequency dependence of the point spread function (PSF) of any interferometer, which can create spectral structure from spectrally smooth foregrounds in our 3D maps (see \cite{EoRWindow1} for a rigorous derivation). Fortunately, in $k_\|$-$k_\perp$ space, instrumental chromaticity from flat-spectrum sources is restricted to the region below 
\beq  
k_\| = \theta_0 \frac{ D_M(z) E(z)}{D_H (1+z)} k_\perp, \label{eq:wedge}
\eeq
where $D_H \equiv c/H_0$, $E(z) \equiv \sqrt{\Omega_m(1+z)^3+\Omega_\Lambda}$, and $D_m(z) \equiv \int_0^z \mathrm{d}z'/E(z')$ with cosmological parameters from \cite{PlanckCosmoParams}. The size of the region is determined by $\theta_0$, the angle from zenith beyond which foregrounds do not significantly contribute. While most of the foreground emission we observe should appear inside the main lobe of the primary beam, foreground contamination from sources in the sidelobes are also significant compared to the signal \cite{PoberSidelobe,NithyaPitchfork,NithyaPitchfork}. A conservative choice of $\theta_0$ is therefore $\pi/2$, which reflects the fact that the maximum possible delay a baseline can measure corresponds to a source at the horizon \cite{AaronDelay}. Still, this foreground isolation is not foolproof and can be easily corrupted by miscalibration and imperfect data reduction. Further, slowly varying spectral modes just outside the wedge are also affected when the foreground residuals have spectral structure beyond that imprinted by the chromaticity of the interferometer.

To confidently detect the 21\,cm EoR power spectrum, we need rigorous statistical techniques that incorporate models of the cosmological signal, the foregrounds, the instrument, the instrumental noise, and the exact mapmaking procedure. With this information, one may use estimators that preserve as much cosmological information as possible and thoroughly propagate errors due to noise and foregrounds through the analysis pipeline.

The development of such statistical techniques has progressed rapidly over the past few years. The quadratic estimator formalism was adapted \cite{LT11} from previous work on the cosmic microwave background \cite{Maxpowerspeclossless} and galaxy surveys \cite{Maxgalaxysurvey1}. It was accelerated to meet the data volume challenges of 21\,cm tomography \cite{DillonFast} and refined to overcome some of the difficulties of working with real data \cite{X13}.  Further, recent work has shown how to rigorously incorporate the interferometric effects that create the wedge \cite{EoRWindow1,EoRWindow2,mapmaking}, though they rely on precision instrument modeling, including exact per-frequency and per-antenna primary beams and complex gains. A similar technique designed for drift-scanning telescopes using spherical harmonic modes was developed in \cite{Richard,ShawCoaxing}, which also demonstrated the need for a precise understanding of one's instrument.

However, at this early stage in the development of 21\,cm cosmology, precision instrument characterization remains an active area of research \cite{sutinjo2014,AbrahamOrbcomm,NewburghCHIMECal,JonniePrimaryBeam}. We thus pursue a more cautious approach to foreground modeling that reflects our incomplete knowledge of the instrument by modeling the residual foreground covariance from the data itself. As we will show, this mitigates systematics such as calibration errors that would otherwise impart spectral structure onto the foregrounds, corrupting the EoR window. While not a fully Bayesian approach like those of \cite{SutterBayesianImaging} and \cite{GibbsPSE}, our technique discovers both the statistics of the foregrounds and the power spectrum from the data. Our foreground models are subject to certain prior assumptions but are allowed to be data motivated in a restricted space. However, by working in the context of the quadratic estimator formalism, we can benefit from the computational speedups of \cite{DillonFast}. This work is meant to build on those techniques and make them more easily applied to real and imperfect data.

This paper is organized into two main parts. In Section \ref{sec:methods} we discuss the problem of covariance modeling in the context of power spectrum estimation and present a method for the empirical estimation of that foreground model, using MWA data to illustrate the procedure. Then, in Section \ref{sec:results}, we explain how these data were taken and reduced into maps and present the results of our power spectrum estimation procedure on a few hours of MWA observation, including limits on the 21\,cm power spectrum.


\section{Empirical Covariance Modeling} \label{sec:methods}

Before presenting our method of empirically modeling the statistics of residual foregrounds in our maps, we need to review the importance of these covariances to power spectrum estimation. We begin in Section \ref{sec:review} with a brief review of the quadratic estimator formalism for optimal power spectrum estimation and rigorous error quantification. We then discuss in Section \ref{sec:CovTheory} the problem of covariance modeling in greater detail, highlighting exactly which unknowns we are trying to model with the data. Next we present in Section \ref{sec:empirical} our empirical method of estimating the covariance of foreground residuals, illustrated with an application to MWA data. Lastly, we review in Section \ref{sec:caveats} the assumptions and caveats that we make or inherit from previous power spectrum estimation work.


\subsection{Quadratic Power Spectrum Estimator Review}\label{sec:review}

The fundamental goal of power spectrum estimation is to reduce the volume of data by exploiting statistical symmetries while retaining as much information as possible about the cosmological power spectrum \cite{Maxpowerspeclossless}. We seek to estimate a set of band powers $\p$ using the approximation that
\beq
P(\kvec) \approx \sum_\alpha p_\alpha \chi_\alpha (\kvec),
\eeq
where $P(\kvec)$ is the power spectrum as a function of wave vector $\kvec$ and $\chi_\alpha$ is an indicator function that equals 1 wherever we are approximating $P(\kvec)$ by $p_\alpha$ and vanishes elsewhere.

Following \cite{LT11,DillonFast,X13}, we estimate power spectra from a ``data cube''---a set of sky maps of brightness temperature at many closely spaced frequencies---which we represent as a single vector $\xhat$ whose index iterates over both position and frequency. From $\xhat$, we estimate each band power as
\beq
\widehat{p}_\alpha = \frac{1}{2}M_{\alpha\beta} \left(\xhat_1 - \mean\right)^\trans \C^{-1} \mathbf{C},_\beta \C^{-1} \left(\xhat_2 -\mean\right) - b_\alpha. \label{eq:QuadEst}
\eeq
Here $\mean = \langle \xhat \rangle$, the ensemble average of our map over many different realizations of the observation, and $\C$ is the covariance of our map,
\beq
\C = \langle \xhat \xhat^\trans \rangle - \langle \xhat \rangle  \langle \xhat \rangle^\trans.
\eeq 
$\mathbf{C},_\beta$ is a matrix that encodes the response of the covariance to changes in the true, underlying band powers; roughly speaking, it performs the Fourier transforming, squaring, and binning steps one normally associates with computing power spectra.\footnote{For a derivation of an explicit form of $\C,_\beta$, see \cite{LT11} or \cite{DillonFast}.}  Additionally, $\M$ is an invertible normalization matrix and $b_\alpha$ is the power spectrum bias from nonsignal contaminants in $\xhat$. In this work, we follow \cite{X13} and choose a form of $\M$ such that $\boldsymbol\Sigma \equiv \text{Cov}(\widehat{\p})$ is diagonal, decorrelating errors in the power spectrum and thus reducing foreground leakage into the EoR window. In order to calculate $\M$ and $\boldsymbol\Sigma$ quickly, we use the fast method of \cite{DillonFast} which uses fast Fourier transforms and Monte Carlo simulations to approximate these matrices.

Finally, temporally interleaving the input data into two cubes $\xhat_1$ and $\xhat_2$ with the same sky signal but independent noise avoids a noise contribution to the bias $b_\alpha$ as in \cite{X13}. Again following \cite{X13}, we abstain from subtracting a foreground residual bias in order to avoid any signal loss (as discussed in \ref{sec:filter}).


\subsection{What Does Our Covariance Model Represent?} \label{sec:CovTheory}

Our brightness temperature data cubes are made up of contributions from three statistically independent sources: the cosmological signal, $\xhat^S$; the astrophysical foregrounds, $\xhat^{FG}$; and the instrumental noise $\xhat^N$. It follows that the covariance matrix is equal to the sum of their separate covariances:
\beq
\C = \C^S + \C^{FG} + \C^N.
\eeq

Hidden in the statistical description of the different contributions to our measurement is an important subtlety. Each of these components is taken to be a particular instantiation of a random process, described by a mean and covariance. In the case of the cosmological signal, it is the underlying statistics---the mean and covariance---which encode information about the cosmology and astrophysics. However, we can only learn about those statistics by assuming statistical isotropy and homogeneity and by assuming that spatial averages can stand in for ensemble averages in large volumes. In the case of the instrumental noise, we usually think of the particular instantiation of the noise that we see as the result of a random trial. 

The foregrounds are different. There is only one set of foregrounds, and they are not random. If we knew exactly how the foregrounds appear in our observations, we would subtract them from our maps and then ignore them in this analysis. We know that we do not know the foregrounds exactly, and so we choose to model them with our best guess, $\mean^{FG}$. If we define the cosmological signal to consist only of fluctuations from the brightness temperature of the global 21\,cm signal, then the signal and the noise both have $\mean^S = \mean^N = 0$. Therefore, we start our power spectrum estimation using Equation \eqref{eq:QuadEst} by subtracting off our best guess as to the foreground contamination in our map. But how wrong are we? 

The short answer is that we do not really know that either. But, if we want to take advantage of the quadratic estimator formalism to give the highest weight to the modes we are most confident in, then we must model the statistics of our foreground residuals. If we assume that our error is drawn from some correlated Gaussian distribution, then we should use that \emph{foreground uncertainty covariance} as the proper $\C^{FG}$ in Equation \eqref{eq:QuadEst}.

So what do we know about the residual foregrounds in our maps? In theory, our dirty maps are related to the true sky by a set of point spread functions that depend on both position and frequency \cite{mapmaking}. This is the result of both the way our interferometer turns the sky into measured visibilities and the way we make maps to turn those visibilities into $\xhat$. In other words, there exists some matrix of PSFs, $\PSF$ such that
\beq
\langle \xhat \rangle = \PSF \x^\text{true}.
\eeq
The spectral structure in our maps that creates the wedge feature in the power spectrum is a result of $\PSF$.

We can describe our uncertainty about the true sky---about the positions, fluxes, and spectral indices of both diffuse foregrounds and points sources---with a covariance matrix $\C^{FG,\text{true}}$ \cite{LT11,DillonFast}, so that
\beq
\C^{FG} = \PSF \C^{FG,\text{true}} \PSF^\trans.
\eeq
This equation presents us with two ways of modeling the foregrounds. If we feel that we know the relationship between our dirty maps and the true sky precisely, then we can propagate our uncertainty about a relatively small number of foreground parameters, as discussed by \cite{LT11} and \cite{DillonFast}, through the $\PSF$ matrix to get $\C^{FG}$. This technique, suggested by \cite{mapmaking}, relies on precise knowledge of $\PSF$. Of course, the relationship between the true sky and our visibility data depends both on the design of our instrument and on its calibration. If our calibration is very good---if we really understand our antenna gains and phases, our primary beams, and our bandpasses---then we can accurately model $\PSF$.

If we are worried about systematics (and at this early stage of 21\,cm tomography with low frequency radio interferometers, we certainly are), then we need a complementary approach to modeling $\C^{FG}$ directly, one that we can use both for power spectrum estimation and for comparison to the results of a more theoretically motivated technique. This is the main goal of this work.


\subsection{Empirical Covariance Modeling Technique} \label{sec:empirical}

The idea of using empirically motivated covariance matrices in the quadratic estimator formalism has some history in the field. Previous MWA power spectrum analysis \cite{X13} used the difference between time-interleaved data cubes to estimate the overall level of noise, empirically calibrating $T_\text{sys}$, the system temperature of the elements. PAPER's power spectrum analysis relies on using observed covariances to suppress systematic errors \cite{PAPER32Limits} and on boot-strapped error bars \cite{PAPER32Limits,DannyMultiRedshift}. A similar technique was developed contemporaneously with this work and was used by \cite{PAPER64Limits} to estimate covariances.

$\C^{FG}$ has far more elements than we have measured voxels---our cubes have about $2\times 10^5$ voxels, meaning that $\C^{FG}$ has up to $2\times 10^{10}$ unique elements. Therefore, any estimate of $\C^{FG}$ from the data needs to make some assumptions about the structure of the covariance. Since foregrounds have intrinsically smooth spectra, and since one generally attempts to model and subtract smooth spectrum foregrounds, it follows that foreground residuals will be highly correlated along the line of sight. After all, if we are undersubtracting foregrounds at one frequency, we are probably undersubtracting at nearby frequencies too. We therefore choose to focus on empirically constructing the part of $\C^{FG}$ that corresponds to the frequency-frequency covariance---the covariance along the line of sight.  If there are $n_f$ frequency channels, then that covariance matrix is only $n_f \times n_f$ elements and is likely dominated by a relatively small number of modes. 

In this section, we will present an approach to solving this problem in a way that faithfully reflects the complex spectral structure introduced by an (imperfectly calibrated) interferometer on the bright astrophysical foregrounds. As a worked example, we use data from a short observation with the MWA which we will describe in detail in Section \ref{sec:results}. We begin with a uniformly weighted map of the sky at each frequency, a model for both point sources and diffuse emission imaged from simulated visibilities, and a model for the noise in each $uv$ cell as a function of frequency.

The idea to model $\C^{FG}$ empirically was put forward by \citet{LiuThesis}. He attempted to model each line of sight as statistically independent and made no effort to separate $\C^{FG}$ from $\C^{N}$ or to reduce the residual noisiness of the frequency-frequency covariance.  

Our approach centers on the idea that the covariance matrix can be approximated as block diagonal in the $uv$ basis of Fourier modes perpendicular to the line of sight. In other words, we are looking to express $\C^{FG}$ as
\beq
C^{FG}_{uu'vv'ff'} \approx \delta_{uu'} \delta_{vv'} \widehat{C}_{ff'}(k_\perp), \label{eq:blockdiag}
\eeq
where $k_\perp$ is a function of $\sqrt{u^2+v^2}$. This is the tensor product of our best guess of the frequency-frequency covariance $\CHat$ and the identity in both Fourier coordinates perpendicular to the line of sight. In this way, we can model different frequency-frequency covariances as a function of $|u|$ or equivalently, $k_\perp$, reflecting that fact that the wedge results from greater leakage of power up from low $k_{\|}$ as one goes to higher $k_\perp$. This method also has the advantage that $\C$ becomes efficient to both write down and invert directly, removing the need for the preconditioned conjugate gradient algorithm employed by \cite{DillonFast}.

This approximation is equivalent to the assumption that the residuals in every line of sight are statistically independent of position. This is generally a pretty accurate assumption as long as the primary beam does not change very much over the map from which we estimate the power spectrum. However, because $\widehat{C}_{ff'}(k_\perp)$ depends on the angular scale, we are still modeling correlations that depend only on the distance between points in the map. 

While we might expect that the largest residual voxels correspond to errors in subtracting the brightest sources, the voxels in the residual data cube (the map minus the model) are only weakly correlated with the best-guess model of the foregrounds (we find a correlation coefficient $\rho =$ \modelResidualCorr, which suggests that sources are removed to roughly the 10\% level, assuming that undersubtraction dominates). As we improve our best guess of the model foregrounds through better deconvolution, we expect $\rho$ to go down, improving the assumption that foregrounds are block diagonal in the $uv$ basis. We will now present the technique we have devised in four steps, employing MWA data as a method demonstration.


\subsubsection{Compute sample covariances in $uv$ annuli}

We begin our empirical covariance calculation by taking the residual data cubes, defined as
\beq
\xhat^\text{res} \equiv \xhat_1/2 + \xhat_2/2 - \mean,
\eeq
and performing a discrete Fourier transform\footnote{For simplicity, we used the unitary discrete Fourier transform for these calculations and ignore any factors of length or inverse length that might come into these calculations only to be canceled out at a later step.} at each frequency independently to get $\widetilde{\mathbf{x}}^\text{res}$. This yields $n_x\times n_y$ sample ``lines of sight'' ($uv$ cells for all frequencies), as many as we have pixels in the map. As a first step toward estimating $\CHat$, we use the unbiased sample covariance estimator from these residual lines of sight. However, instead of calculating a single frequency-frequency covariance, we want to calculate many different $\CHat^\text{res}$ matrices to reflect the evolution of spectral structure with $k_\perp$ along the wedge. We therefore break the $uv$ plane into concentric annuli of equal width and calculate $\CHat^\text{res}_{uv}$ for each $uv$ cell as the sample covariance of the $N^\text{LOS} - 2$ lines of sight in that annulus, excluding the cell considered and its complex conjugate. Since the covariance is assumed to be block diagonal, this eliminates a potential bias that comes from downweighting a uv cell using information about that cell. Thus,
\beq
\widehat{C}_{uv,ff'}^\text{res} = \hspace{-3mm}\sum_{\substack{\text{other } u',v' \\ \text{in annulus}}} \hspace{-3mm} \frac{\left(\widetilde{x}^\text{res}_{u'v'f} - \langle \widetilde{x}^\text{res}_f \rangle\right) \left(\widetilde{x}^\text{res}_{u'v'f'} - \langle \widetilde{x}^\text{res}_{f'} \rangle \right)^*}{N^\text{LOS}-2-1}, \label{eq:CovEstOtherAnnuli}
\eeq
where $\langle \widetilde{x}^\text{res}_f \rangle$ is an average over all $u'$ and $v'$ in the annulus. We expect this procedure to be particularly effective in our case because the $uv$ coverage of the MWA after rotation synthesis is relatively symmetric.

\begin{figure}[]  
	\centering 
	\includegraphics[width=.48\textwidth]{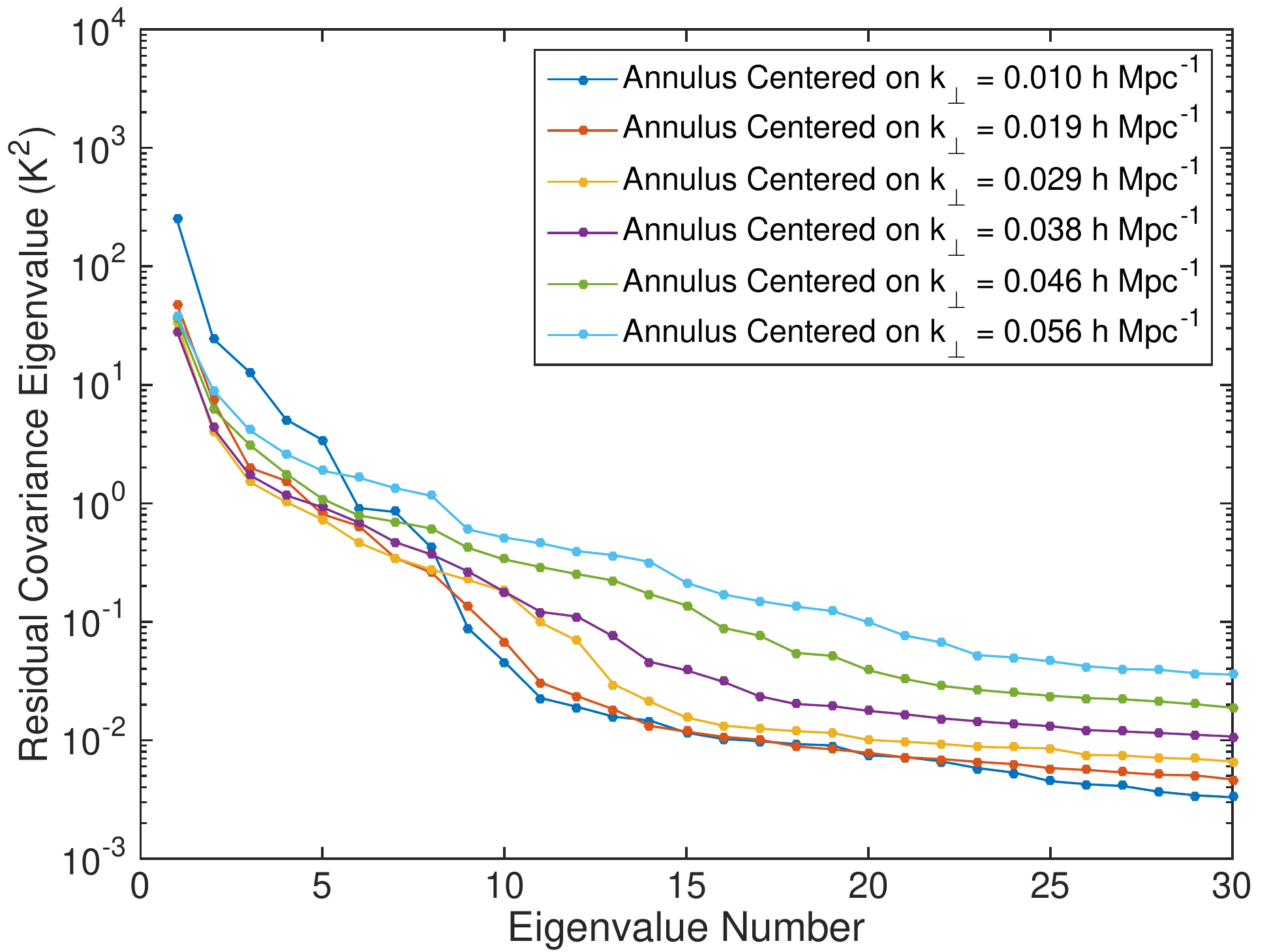}
	\caption{The evolution of the wedge with $k_\perp$ motivates us to model foregrounds separately for discrete values of $k_\perp$. In this plot of the 30 largest eigenvalues of the observed residual covariance (which should include both noise and foregrounds) sampled in six concentric annuli, we see steeper declines toward a noise floor for the inner annuli than the outer annuli. This is consistent with the expected effect of the wedge---higher $k_\perp$ modes should be foreground contaminated at higher $k_\|$.}
	\label{fig:wedge_eigenvalues}
\end{figure}

As a sense check on these covariances, we plot their largest 30 eigenvalues in Figure \ref{fig:wedge_eigenvalues}. We see that as $|u|$ (and thus $k_\perp$) increases, the eigenspectra become shallower. At high $k_\perp$, the effect of the wedge is to leak power to a range of $k_\|$ values. The eigenspectrum of intrinsically smooth foregrounds should be declining exponentially \cite{AdrianForegrounds}. The wedge softens that decline. These trends are in line with our expectations and further motivate our strategy of forming covariance matrices for each annulus independently. 

Because we seek only to estimate the foreground portion of the covariance, the formal rank deficiency of $\CHat^\text{res}_{uv}$ is not a problem.\footnote{In fact, the rank of $\CHat^\text{res}_{uv}$ is $N_\text{LOS} - 3$ if $N^\text{LOS}-2 \le n_f$.} All we require is that the largest (and thus more foreground-dominated) modes be well measured. In this analysis, we used six concentric annuli to create six different frequency-frequency foreground covariances. Using more annuli allows for better modeling of the evolution of the wedge with $k_\perp$ at the expense of each estimate being more susceptible to noise and rank deficiency. 


\subsubsection{Subtract the properly projected noise covariance.}
 
The covariances computed from these $uv$ lines of sight include contributions from the 21\,cm signal and instrumental noise as well as foregrounds. We can safely ignore the signal covariance for now as we are far from the regime where sample variance is significant. We already have a theoretically motivated model for the noise (based on the $uv$ sampling) that has been empirically rescaled to the observed noise in the difference of time-interleaved data (the same basic procedure as in \cite{X13}). We would like an empirical estimate of the residual foreground covariance alone to use in $\C^{FG}$ and thus must subtract off the part of our measurement we think is due to noise variance.

To get to $\CHat^\text{FG}_{uv}$ from $\CHat^\text{res}_{uv}$, we subtract our best guess of the portion of $\CHat^\text{res}_{uv}$ that is due to noise, which we approximate by averaging the noise model variances in all the other $uv$ cells in the annulus at that given frequency, yielding
\beq
\widehat{C}_{uv,ff'}^\text{N} = \frac{1}{N^\text{LOS}} \hspace{-2mm}  \sum_{\substack{\text{other } u',v' \\ \text{in annulus}}}  \hspace{-2mm} \delta_{uu'} \delta_{vv'} \delta_{ff'} C^{N}_{uu'vv'ff'}.
\eeq
Note, however, that $\CHat_{uv}^\text{N}$ is full rank while $\CHat_{uv}^{res}$ is typically rank deficient. Thus a naive subtraction would oversubtract the noise variance in the part of the subspace of $\CHat_{uv}^\text{N}$ where $\CHat_{uv}^{res}$ is identically zero. Instead, the proper procedure is to find the projection matrices $\Proj_{uv}$ that discard all eigenmodes outside the subspace where $\CHat_{uv}^\text{res}$ is full rank. Each should have eigenvalues equal to zero or one only and have the property that
\beq
\Proj_{uv} \CHat_{uv}^\text{res} \Proj_{uv}^\trans = \CHat_{uv}^\text{res}.
\eeq
Only after projecting out the part of $\CHat^N_{uv}$ inside the unsampled subspace can we self-consistently subtract our best guess of the noise contribution to the subspace in which we seek to estimate foregrounds. In other words, we estimate $\CHat^\text{FG}_{uv}$ as
\beq
\CHat^\text{FG}_{uv} = \CHat_{uv}^\text{res} - \Proj_{uv} \CHat_{uv}^\text{N} \Proj_{uv}^\trans.
\eeq

\begin{figure*}[]  
	\centering 
	\includegraphics[width=1\textwidth]{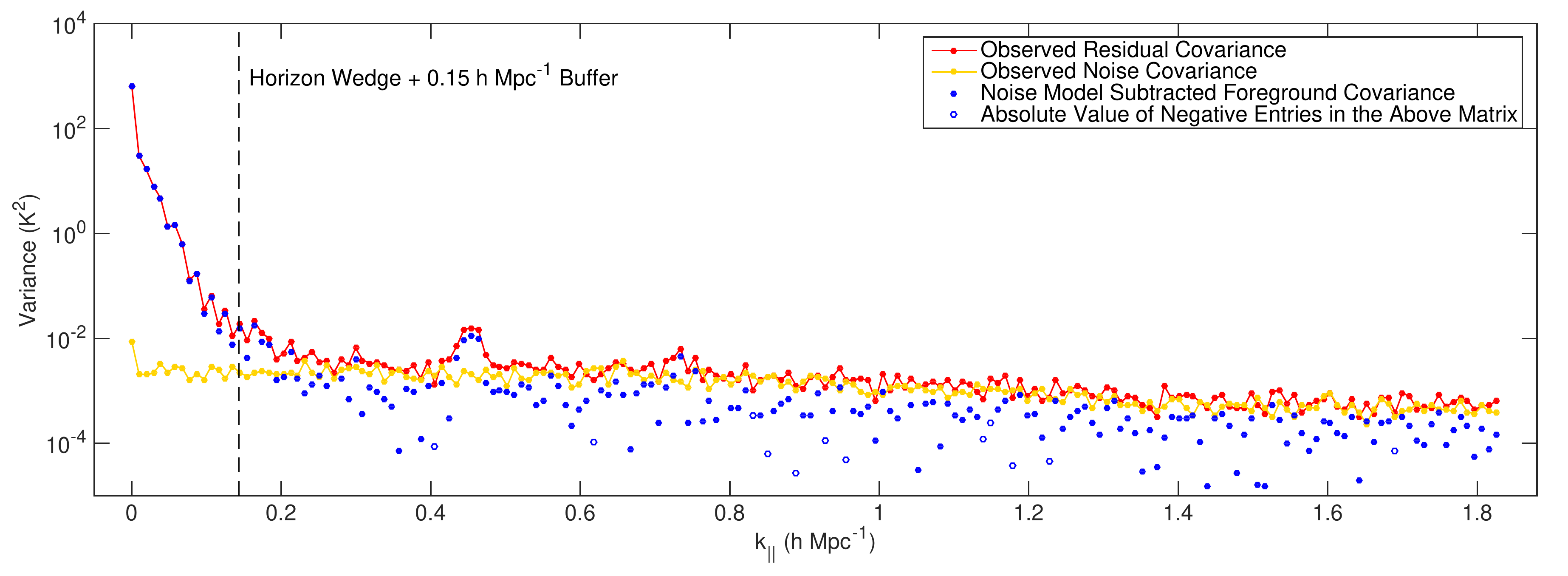}
	\caption{Examining the diagonal elements of the observed residual and inferred foreground covariance matrices in Fourier space reveals the effectiveness of subtracting model for the noise covariance. In red, we plot the observed residual covariance, which contains both foregrounds and noise. As a function of $k_\|$, the two separate relatively cleanly---there is a steeply declining foreground portion on the left followed by a relatively flat noise floor on the right. The theory that the right-hand portion is dominated by noise is borne out by the fact that it so closely matches the observed noise covariance, inferred lines of sight of $\x_1 - \x_2$, which should have only noise and no sky signal at all. The regions where they differ significantly, for example at $k_\| \sim 0.45$\,$h$\,Mpc$^{-1}$ , are attributable to systematic effects like the MWA's coarse band structure that have not been perfectly calibrated out. For the example covariances shown here (which correspond to a mode in the annulus at $k_\perp \approx 0.010$\,$h$\,Mpc$^{-1}$), we can see that subtracting a properly projected noise covariance removes most of the power from the noise-dominated region, leaving only residual noise that appears both as negative power (open blue circle) and as positive power (closed blue circles) at considerably lower magnitude.}
	\label{fig:fourier_diags}
\end{figure*}

We demonstrate the effectiveness of this technique in Figure \ref{fig:fourier_diags} by plotting the diagonal elements of the Fourier transform of $\CHat_{uv}^\text{res}$ and $\CHat_{uv}^\text{FG}$ along the line of sight. Subtracting of the noise covariance indeed eliminates the majority of the power in the noise dominated modes at high $k_\|$; thus we expect it also to fare well in the transition region near the edge of the wedge where foreground and noise contributions are comparable.

\subsubsection{Perform a $k_\|$ filter on the covariance.} \label{sec:filter}

\begin{figure}[] 
	\centering 
	\includegraphics[width=.48\textwidth]{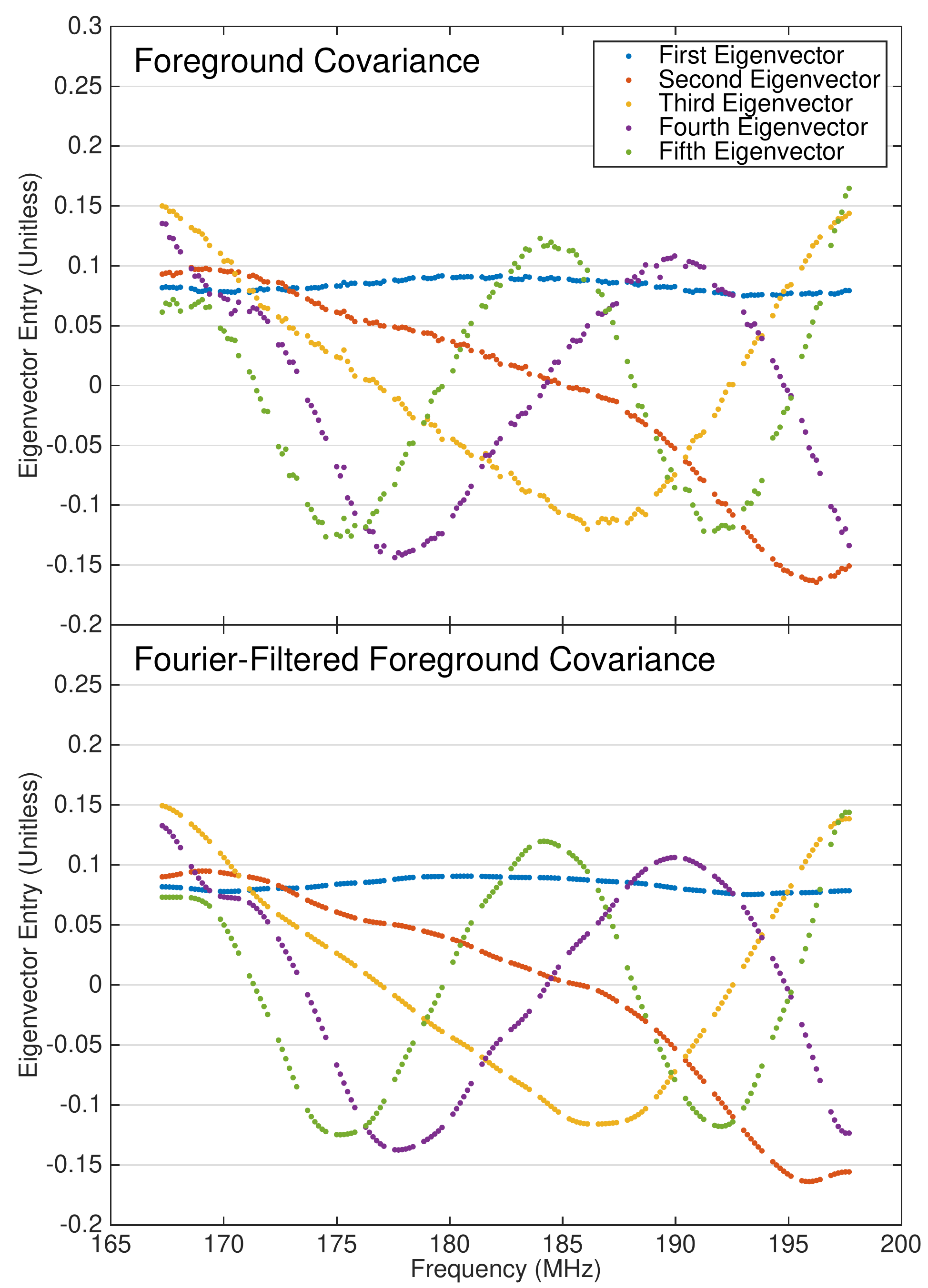} 
	\caption{The foreground covariance we estimate from our limited data set is still very noisy, and we run the risk of overfitting the noise in our measurements if we take it at face value. In the top panel, we plot the eigenvectors corresponding to the five largest eigenvalues of $\CHat^\text{FG}$ for a mode in the annulus centered on $k_\perp \approx 0.010$\,$h$\,Mpc$^{-1}$.  In the bottom panel, we show dominant eigenvectors of the Fourier-filtered covariance. As expected, they resemble the first five Fourier modes. The missing data every 1.28 MHz are due to channels flagged at the edge of the coarse bandpass of the MWA's polyphase filter bank---the most difficult part of the band to calibrate.} 
	\label{fig:eigenvectors}
\end{figure}

Despite the relatively clean separation of foreground and noise eigenvalues, inspection of some of the foreground-dominated modes in the top panel of Figure \ref{fig:eigenvectors} reveals residual noise. Using a foreground covariance constructed from these noisy foreground eigenmodes to downweight the data during power spectrum estimation would errantly downweight some high $k_\|$ modes in addition to the low $k_\|$ foreground-dominated modes. To avoid this double counting of the noise, we allow the foreground covariance to include only certain $k_\|$ modes by filtering $\CHat^\text{FG}_{uv}$ in Fourier space to get $\CHat^\text{FG,filtered}_{uv}$. Put another way, we are imposing a prior on which Fourier modes we think have foreground power in them. The resulting noise filtered eigenmodes are shown in the bottom panel of Figure \ref{fig:eigenvectors}.

In practice, implementing this filter is subtle. We interpolate $\CHat^\text{FG}$ over the flagged frequency channels using a cubic spline, then symmetrically pad the covariance matrix, forcing its boundary condition to be periodic. We then Fourier transform, filter, inverse Fourier transform, remove the padding, and then rezero the flagged channels. 

Selecting a filter to use is also a subtle choice. We first keep modes inside the horizon wedge with an added buffer. For each annulus, we calculate a mean value of $k_\perp$, and then use Equation \eqref{eq:wedge} to calculate the $k_\|$ value of the horizon wedge, using $\theta_0 = \pi/2$. Although the literature suggests a $0.1$ to $0.15$\,$h$\,Mpc$^{-1}$ buffer for ``suprahorizon emission'' due to some combination of intrinsic spectral structure of foregrounds, primary beam chromaticity, and finite bandwidth \cite{PoberWedge,PoberNextGen}, we pick a conservative $0.5$\,$h$\,Mpc$^{-1}$. Then we examine the diagonal of $\CHat^\text{FG}$ (Figure \ref{fig:fourier_diags}) to identify additional foreground modes, this time in the EoR window, due to imperfect bandpass calibration appearing as spikes. One example is the peak at $k_\| \sim 0.45$\,$h$\,Mpc$^{-1}$. Such modes contribute errant power to the EoR window at constant $k_\|$. Since these modes result from the convolution of the foregrounds with our instrument, they also should be modeled in $\C^{FG}$ in order to minimize their leakage into the rest of the EoR window. 

One might be concerned that cosmological signal and foregrounds theoretically both appear in the estimate of $\C^{FG}$ that we have constructed, especially with our conservative $0.5$\,$h$\,Mpc$^{-1}$ buffer that allows foregrounds to be discovered well into the EoR window. For the purposes of calculating $\C^{-1}(\widehat{\x}-\mean)$ in the quadratic estimator in Equation \eqref{eq:QuadEst}, that is fine since its effect is to partially relax the assumption that sample variance can be ignored. However, the calculation of the bias depends on being able to differentiate signal from contaminants \cite{Maxpowerspeclossless,LT11,DillonFast}. 

The noise contribution to the bias can be eliminated by cross-correlating maps made from interleaved time steps \cite{X13}. However, we cannot use our inferred $\C^{FG}$ to subtract a foreground bias without signal loss. That said, we can still set an upper limit on the 21\,cm signal. By following the data and allowing the foreground covariance to have power inside the EoR window, we are minimizing the leakage of foregrounds into uncontaminated regions and we are accurately marking those regions as having high variance. As calibration and the control of systematic effects improves, we should be able to isolate foregrounds to outside the EoR window, impose a more aggressive Fourier filter on $\C^{FG}$, and make a detection of the 21\,cm signal by employing foreground avoidance.


\subsubsection{Cut out modes attributable to noise.}

After suppressing the noisiest modes with our Fourier filter, we must select a cutoff beyond which the foreground modes are irrecoverably buried under noise. We do this by inspecting the eigenspectrum of $\CHat^\text{FG,filtered}_{uv}$. The true $\C^{FG}$, by definition, admits only positive eigenvalues (though some of them should be vanishingly small). 

By limiting the number of eigenvalues and eigenvectors we ultimately associate with foregrounds, we also limit the potential for signal loss by allowing a large portion of the free parameters to get absorbed into the contaminant model \cite{EricAdrianEmpiricalGlobalForegrounds,PAPER64Limits}. When measuring the power spectrum inside the EoR window, we can be confident that signal loss is minimal compared to foreground bias and other errors.

\begin{figure}[]  
	\centering 
	\includegraphics[width=.48\textwidth]{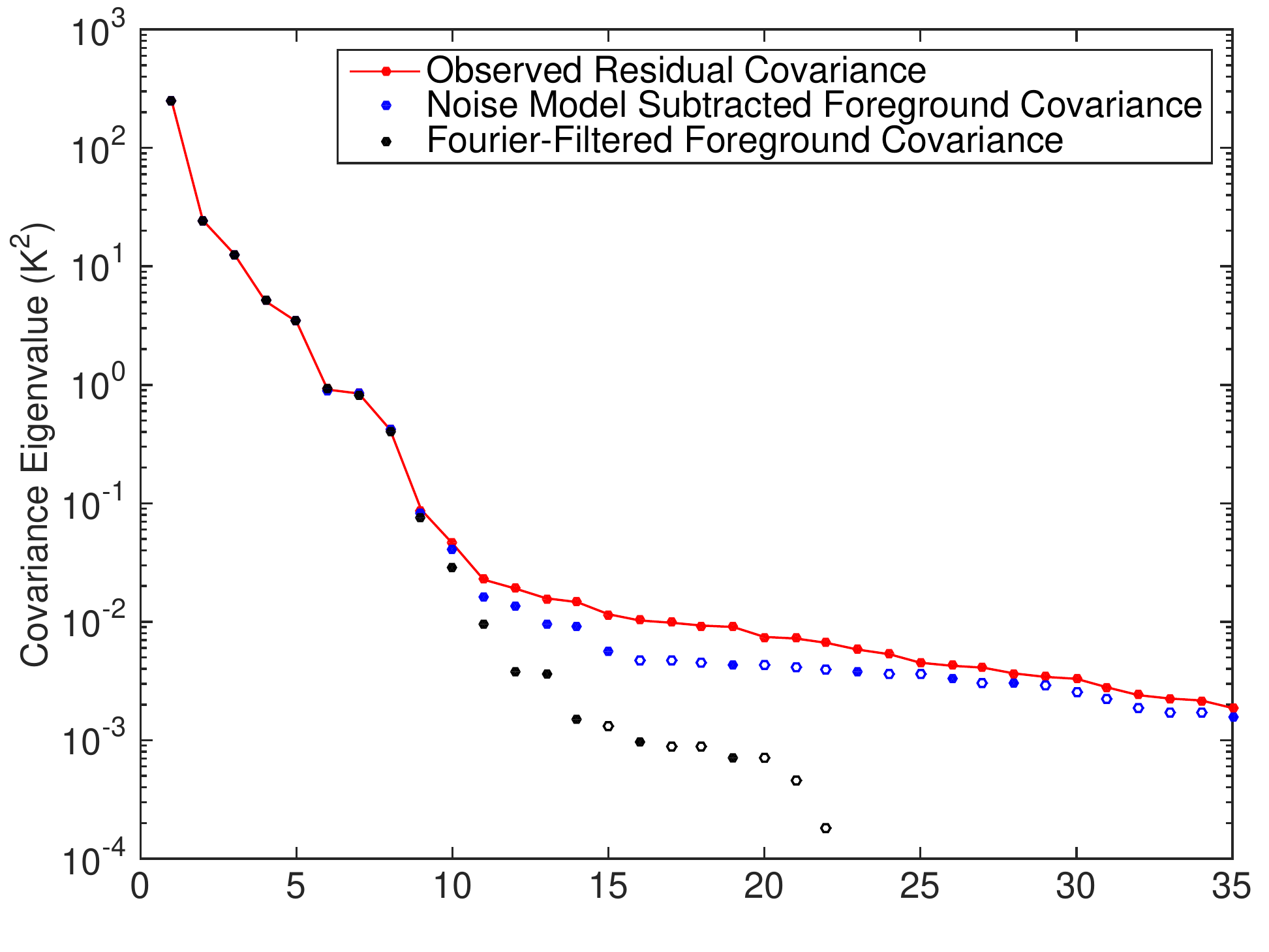}
	\caption{The evolution of the eigenvalues of our estimated foreground covariance matrix for a mode in the annulus corresponding to $k_\perp \approx 0.010$\,$h$\,Mpc$^{-1}$ at each of the first three stages of covariance estimation. First we calculate a sample covariance matrix from the residual data cubes (shown in red). Next we subtract our best guess as to the part of the diagonal of that matrix that originates from instrumental noise, leaving the blue dots (open circles are absolute values of negative eigenvalues). Then we filter out modes in Fourier space along the line of sight that we think should be noise dominated, leaving the black dots. Finally, we project out the eigenvectors associated with eigenvalues whose magnitude is smaller than the largest negative eigenvalue, since those are likely due to residual noise. What remains is our best guess at the foreground covariance in an annulus and incorporates as well as possible our prior beliefs about its structure.}
	\label{fig:cov_steps}
\end{figure}

We plot in Figure \ref{fig:cov_steps} the eigenspectra of $\CHat_{uv}^\text{res}$, $\CHat^\text{FG}_{uv}$, and $\CHat^\text{FG,filtered}_{uv}$, sorted by absolute value. There are two distinct regions---the sharply declining foreground-dominated region and a flatter region with many negative eigenvalues. We excise eigenvectors whose eigenvalues are smaller in absolute value than the most negative eigenvalue. This incurs a slight risk of retaining a few noise dominated modes, albeit strongly suppressed by our noise variance subtraction and our Fourier filtering. Finally we are able to construct the full covariance $\CHat$ using Equation \eqref{eq:blockdiag}.

\subsection{Review of Assumptions and Caveats} \label{sec:caveats}

Before proceeding to demonstrate the effectiveness of our empirical covariance modeling method, it is useful to review and summarize the assumptions made about mapmaking and covariance modeling. Some are inherited from the previous application of quadratic power spectrum estimation to the MWA \cite{X13}, while others are necessitated by our new, more faithful foreground covariance. Relaxing these assumptions in a computationally efficient manner remains a challenge we leave for future work.

\begin{enumerate}[i.]
\item We adopt the flat sky approximation as in \cite{DillonFast,X13}, allowing us to use the fast Fourier transform to quickly compute power spectra. The error incurred from this approximation on the power spectrum is expected to be smaller than 1\% \cite{X13}.

\item We assume the expectation value of our uniformly weighted map is the true sky (i.e., $\langle \widehat{\x} \rangle = \x^\text{true}$) when calculating $\C,_\beta$ in Equation \ref{eq:QuadEst}, again following \cite{X13}. In general $\langle \widehat{\x} \rangle$ is related to $\x^\text{true}$ by $\PSF$, the matrix of point spread functions \cite{mapmaking}. Here we effectively approximate the PSF as position independent. Relaxing this approximation necessitates the full mapmaking theory presented in \cite{mapmaking} which has yet to be integrated into a power spectrum estimation pipeline. 

\item We approximate the foreground covariance as uncorrelated between different $uv$ cells (and thus block diagonal). At some level there likely are correlations in $uv$, though those along the line of sight are far stronger. It may be possible to attempt to calculate these correlations empirically, but it would be very difficult considering relative strength of line-of-sight correlations. It may also be possible to use a nonempirical model, though that has the potential to make the computational speedups of \cite{DillonFast} more difficult to attain.
 
\item We approximate the frequency-frequency foreground covariance as constant within each annulus, estimating our covariance for each $uv$ cell only from other cells in the same annulus. In principle, even if the foreground residuals were isotropic, there should be radial evolution within each annulus which we ignore for this analysis. 

\item The Fourier filter is a nontrivial data analysis choice balancing risk of noise double counting against that of insufficiently aggressive foreground downweighting.

\item In order to detect the 21\,cm signal, we assume that foregrounds can be avoided by working within the EoR window. Out of fear of losing signal, we make no effort to subtract a residual foreground bias from the window. This makes a detection inside the wedge impossible and it risks confusing foreground contamination in the window for a signal. Only analysis of the dependence of the measurement on $z$, $k$, $k_\|$, and $k_\perp$ can distinguish between systematics and the true signal. 

\end{enumerate}


\section{Results} \label{sec:results}

We can now demonstrate the statistical techniques we have motivated and developed in Section \ref{sec:methods} on the problem of estimating power spectra from a 3\,h observation with the 128-antenna MWA. We begin with a discussion of the instrument and the observations in Section \ref{sec:observation}. In Section \ref{sec:processing} we detail the data processing from raw visibilities to calibrated maps from which we estimate both the foreground residual covariance matrix and the power spectrum. Finally, in Section \ref{sec:results} we present our results and discuss lessons learned looking toward a detection of the 21\,cm signal.


\subsection{Observation Summary} \label{sec:observation}

The 128-antenna Murchison Widefield Array began deep EoR observations in mid-2013. We describe here the salient features of the array and refer to \cite{TingaySummary} for a more detailed description. The antennas are laid out over a region of radius 1.5\,km in a quasirandom, centrally concentrated distribution which achieves approximately complete $uv$ coverage at each frequency over several hours of rotation synthesis \cite{MWAsensitivity}. Each antenna element is a phased array of 16 wideband dipole antennas whose phased sum forms a discretely steerable 25$^\circ$ beams (full width at half maximum) at 150\,MHz with frequency-dependent, percent level sidelobes \cite{AbrahamOrbcomm}. We repoint the beam to our field center on a 30\,min cadence to correct for earth rotation, effectively acquiring a series of drift scans over this field. 

We observe the MWA ``EOR0'' deep integration field, centered at R.A.(J2000) $= 0^\text{h}\,0^\text{m}\,0^\text{s}$ and decl.(J2000) $= -30^\circ\,0'\,0''$. It features a near-zenith position, a high Galactic latitude, minimal Galactic emission \cite{GSM}, and an absence of bright extended sources. This last property greatly facilitates calibration in comparison to the ``EOR2'' field---a field dominated by the slightly resolved radio galaxy Hydra A at its center---which was used by \cite{ChrisMWA} and \cite{X13}. A nominal 3\,h set of EOR0 observations was selected during the first weeks of observing to use for refining and comparing data processing, imaging, and power spectra pipelines \cite{JacobsPipelines}. In this work, we use the ``high band,'' near-zenith subset of these observations with 30.72\,MHz of bandwidth and center frequency of 182\,MHz, recorded on Aug 23, 2013 between 16:47:28 and 19:56:32 UTC (22.712 and 1.872 hours LST). 


\subsection{Calibration and Mapmaking Summary} \label{sec:processing}

Preliminary processing, including radio frequency interference (RFI) flagging followed by time and frequency averaging, was performed with the \texttt{COTTER} package \cite{AndreMWARFI} on raw correlator data. These data were collected at 40\,kHz resolution with an integration time of 0.5\,s, and averaged to 80\,kHz resolution with a 2\,s integration time to reduce the data volume. Additionally, 80\,kHz at the upper and lower edges of each of 24 coarse channels (each of width 1.28\,MHz) of the polyphase filter bank is flagged due to known aliasing.

As in \cite{X13}, we undertake snapshot-based processing in which each minute-scale integration is calibrated and imaged independently. Snapshots are combined only at the last step in forming a Stokes $I$ image cube, allowing us to properly align and weight them despite different primary beams due to sky rotation and periodic repointing. 
While sources are forward modeled for calibration and foreground subtraction using the full position dependent PSF (i.e., the synthesized beam), we continue to approximate it as position independent (and equal to that of a point source at the field center) during application of uniform weighting and computation of the noise covariance. 

We use the calibration, foreground modeling, and first stage image products produced by the Fast Holographic Deconvolution\footnote{For a theoretical discussion of the algorithm see \cite{FHD}. The code is available at \url{https://github.com/miguelfmorales/FHD}.} (FHD) pipeline as described by \cite{JacobsPipelines}. The calibration implemented in the FHD package is an adaptation of the fast algorithm presented by \cite{Salvini2014} with a baseline cutoff of $b>50\lambda$. In this data reduction, the point source catalogs discussed below are taken as the sky model for calibration.  Solutions are first obtained per antenna and per frequency before being constrained to linear phase slopes and quadratic amplitude functions after correcting for a median antenna-independent amplitude bandpass.  The foreground model used for subtraction includes models both of diffuse radio emission \cite{AdamDiffuse} and point sources. In detail, the point source catalog is the union of a deep MWA point source survey within $20^\circ$ of the field center \cite{PattiCatalog1}, the shallower but wider MWA commissioning point source survey \cite{MWACS}, and the Culgoora catalog \cite{Slee1995}. Note that calibration and foreground subtraction of off-zenith observations are complicated by Galactic emission picked up by primary beam sidelobes, and are active topics of investigation \cite{PoberSidelobe,NithyaPitchfork,NithyaPitchforkConfirmation}. During these observations a single antenna was flagged due to known hardware problems, and 1--5 more were flagged for any given snapshot due to poor calibration solutions.

These calibration, foreground modeling, and imaging steps constitute notable improvements over \cite{X13}. In that work, the presence of the slightly resolved Hydra A in their EOR2 field likely limited calibration and subtraction fidelity as only a point source sky model was used. In contrast, the EOR0 field analyzed here lacks any such nearby radio sources. Our foreground model contains $\sim 2500$ point sources within the main lobe and several thousand more in the primary beam sidelobes in addition to the aforementioned diffuse map. A last improvement in the imaging is the more frequent interleaving of time steps for the cross power spectrum, which we performed at the integration scale (2\,s) as opposed to the snapshot scale (a few minutes). This ensures that both $\xhat_1$ and $\xhat_2$ have identical sky responses and thus allows us to accurately estimate the noise in the array from difference cubes. Assuming that the system temperature contains both an instrumental noise temperature and a frequency dependent sky noise temperature that scales as $\nu^{-2.55}$, the observed residual root-mean-square brightness temperature is consistent with $T_\text{sys}$ ranging from 450\,K at 167\,MHz to 310\,K at 198\,MHz, in line with expectations \cite{MWAsensitivity}.

As discussed in \cite{JacobsPipelines} and \cite{HazeltonEppsilon}, FHD produces naturally weighted sky, foreground model, ``weights,'' and ``variances'' cubes, as well as beam-squared cubes. All are saved in image space using the HEALPix format \cite{HEALPIX} with $N_\text{side}=1024$. Note that these image cubes are crops of full-sky image cubes to a $16^\circ\times16^\circ$ square field of view, as discussed below. The sky, foreground model, and weights cubes are image space representations of the measured visibilities, model visibilities, and sampling function, respectively, all originally gridded in $uv$ space using the primary beam as the gridding kernel. The variances cube is similar to the weights cube, except the gridding kernel is the {\textit square} of the $uv$ space primary beam. It represents the proper quadrature summation of independent noise in different visibilities when they contribute to the same $uv$ cell, and will ultimately become our diagonal noise covariance model. The FHD cubes from all ninety-four 112\,s snapshots are optimally combined in this ``holographic'' frame in which the true sky is weighted by two factors of the primary beam, as in \cite{X13}.

We perform a series of steps to convert the image cube output of FHD into uniformly weighted Stokes $I$ cubes accompanied by appropriate $uv$ coverage information for our noise model. We first map these data cubes onto a rectilinear grid, invoking the flat sky approximation. We do this by rotating the (RA,Dec) HEALPix coordinates of the EOR0 field to the north pole (0$^\circ$,90$^\circ$), and then projecting and gridding onto the $xy$ plane with $0.2^\circ\times0.2^\circ$ resolution over a $16^\circ\times16^\circ$ square field of view. To reduce the data volume while maintaining cosmological sensitivity, we coarse grid to approximately $0.5^\circ$ resolution by Fourier transforming and cropping these cubes in the $uv$ plane at each frequency. We form a uniformly weighted Stokes I cube $I_{\text{uni}}(\vec{\theta})$ by first summing the XX and YY data cubes, resulting in a naturally weighted, holographic stokes I cube $I_{\text{nat},h}(\vec{\theta})  = I_{XX,h}(\vec{\theta})+I_{YY,h}(\vec{\theta})$. Then we divide out the holographic weights cube $W_h({\vec{\theta}})$ in $uv$ space, which applies uniform weighting and removes one image space factor of the beam, and lastly divide out the second beam factor $B(\vec{\theta})$: $I_{\text{uni}}(\vec{\theta}) = \mathcal{F}^{-1}[\mathcal{F}I_{\text{nat},h}(\vec{\theta})/\mathcal{F}W_h(\vec{\theta})]/B(\vec{\theta})$, where $\mathcal{F}$ represents a Fourier transform and $B(\vec{\theta}) = [B_{XX}^2(\vec{\theta})+B_{YY}^2(\vec{\theta})]^{1/2}$. Consistent treatment of the variances cube requires $uv$ space division of {\it two} factors of the weights cube followed by image space division of {\it two} factors of the beam.

Lastly, we frequency average from 80 kHz to 160 kHz, flagging a single 160 kHz channel the edge of each 1.28\,MHz coarse channel due to polyphase filter bank attenuation and aliasing, which make these channels difficult to reliably calibrate. Following \cite{X13}, we also flag poorly observed $uv$ cells and $uv$ cells whose observation times vary widely between frequencies. In all cases, we formally set the variance in flagged channels and $uv$ cells in $\C^N$ to infinity and use the pseudoinverse to project out flagged modes \cite{X13}. 
 
  
\subsection{Power Spectrum Results}
 
We can now present the results of our method applied to 3\,h of MWA-128T data. We first study cylindrically averaged, two-dimensional (2D) power spectra and their statistics, since they are useful for seeing the effects of foregrounds and systematic errors on the power spectrum. We form these power spectra with the full 30.72\,MHz instrument bandwidth to achieve maximal $k_\|$ resolution.

\begin{figure}[] 
	\centering 
	\includegraphics[width=.48\textwidth]{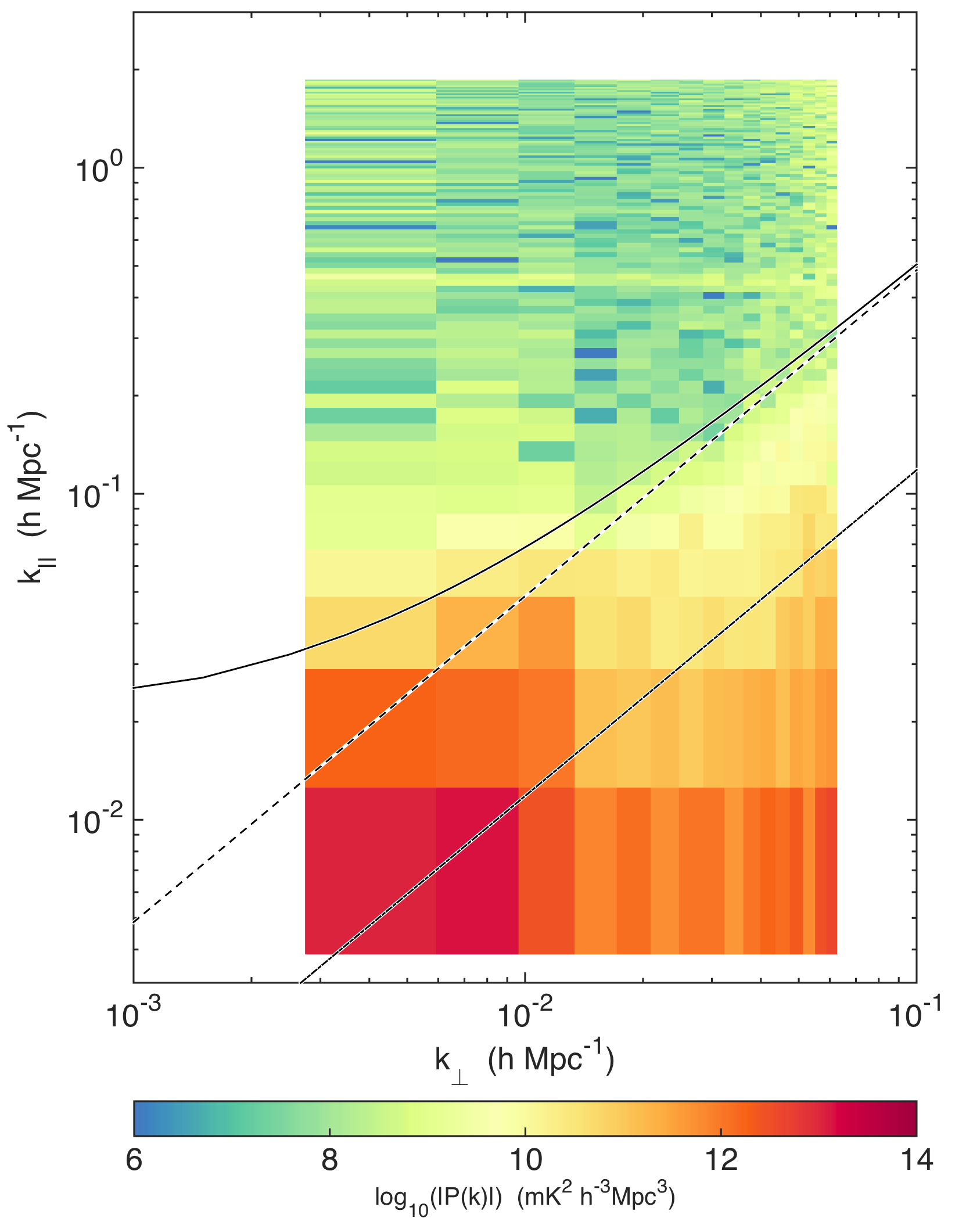}
	\caption{Our power spectrum clearly exhibits the typical EoR window structure with orders-of-magnitude suppression of foregrounds in the EoR window. Here we plot our estimates for $|P(k_\perp,k_\|)|$ for the full instrumental bandwidth, equivalent to the range $z=6.2$ to $z=7.5$. Overplotted is the wedge from Equation \ref{eq:wedge} corresponding to the first null in the primary beam (dash-dotted line), the horizon (dashed line), and the horizon with a relatively aggressive \wedgeBuffer\,$h$\,Mpc$^-1$ buffer (solid curve). In addition to typical foreground structure, we also see the effect of noise at high and very low $k_\perp$ where baseline coverage is poor. We also clearly see a line of power at constant $k_\| \approx 0.45$\,$h$\,Mpc$^{-1}$, attributable to miscalibration of the instrument's bandpass and cable reflections \cite{HazeltonEppsilon}.}
	\label{fig:2dPk}
\end{figure}

We begin with the 2D power spectrum itself (Figure \ref{fig:2dPk}) in which several important features can be observed.
First, the wedge and EoR window are clearly distinguishable, with foregrounds suppressed by at least 5 orders of magnitude across most of the EoR window. At high $k_\perp$, the edge of the wedge is set by the horizon while at low $k_\perp$ the cutoff is less clear. There appears to be some level of suprahorizon emission, which was also observed with PAPER in \cite{PoberWedge} and further explained by \cite{EoRWindow1}. Consistent with Figure \ref{fig:wedge_eigenvalues} we see the strongest foreground residual power at low $k_\perp$, meaning that there still remains a very large contribution from diffuse emission from our Galaxy---potentially from sidelobes of the primary beam affecting the shortest baselines \cite{NithyaPitchfork,NithyaPitchforkConfirmation}.

We also see evidence for less-than-ideal behavior. Through we identified spectral structure appearing at $k_\| \sim 0.45$\,$h$\,Mpc$^{-1}$ in Figure \ref{fig:fourier_diags} and included it in our foreground residual covariance, that contamination still appears here as a horizontal line. By including it in the foreground residual model, we increase the variance we associate with those modes and we decrease the leakage out of those modes, isolating the effect to only a few $k_\|$ bins.

\begin{figure}[] 
	\centering 
	\includegraphics[width=.48\textwidth]{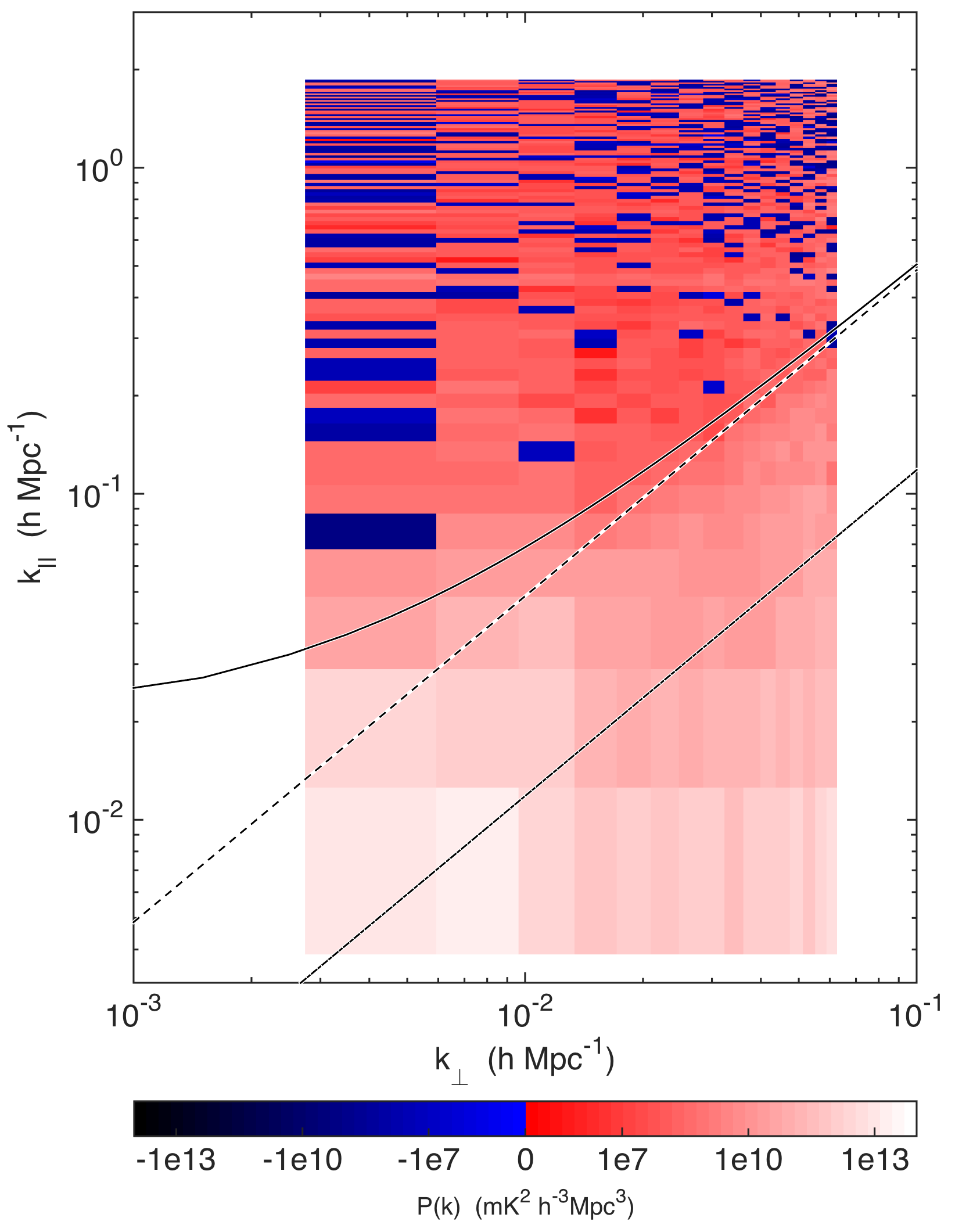}
	\caption{By using an estimator of the power spectrum with uncorrelated errors between bins, we can see that most of the EoR window is noise dominated in our power spectum measurement. Here we show the inverse hyperbolic sine of the power spectum, which behaves linearly near zero and logarithmically at large magnitudes. Because we are taking a cross power spectrum between two data cubes with uncorrelated noise, noise dominated regions are equally likely to have positive power as negative power. Since we do not attempt to subtract a foreground bias, foreground contaminated regions show up as strongly positive. That includes the wedge, the bandpass line at $k_\| \approx 0.45$\,$h$\,Mpc$^{-1}$  (see Figure \ref{fig:2dPk}), and some of the EoR window at low $k_\perp$ and relatively low $k_\|$, consistent with the suprahorizon emission seen in \cite{PoberWedge}.}
	\label{fig:2dPkArcSinh}
\end{figure}

While Figure \ref{fig:2dPk} shows the magnitude of the 2D power spectrum, Figure \ref{fig:2dPkArcSinh} shows its sign using a split color scale, providing another way to assess foreground contamination in the EoR window. 
Because we are taking the cross power spectrum between two cubes with identical sky signal but independent noise realizations, the noise dominated regions should be positive or negative with equal probability. This is made possible by our use of a power spectrum estimator normalized such that $\boldsymbol\Sigma \equiv \text{Cov}(\widehat{\p})$ is a diagonal matrix \cite{DillonFast}. This choice limits leakage of foreground residuals from the wedge into the EoR window \cite{X13}.

By this metric, the EoR window is observed to be noise dominated with only two notable exceptions. The first is the region just outside the wedge at low $k_\perp$ attributable to suprahorizon emission due to some combination of  intrinsic foreground spectral structure, beam chromaticity, and finite bandwidth. This suggests our aggressive 0.02\,$h$\,Mpc$^{-1}$ cut beyond the horizon will leave in some foreground contamination when we bin to form one-dimensional (1D) power spectra. As long as we are only claiming an upper limit on the power spectrum, this is fine. A detection of foregrounds is also an upper limit on the cosmological signal. More subtle is the line of positive power at $k_\| \sim 0.45$\,$h$\,Mpc$^{-1}$, confirming our hypothesis that the spike observed in Figure \ref{fig:2dPk} is indeed an instrumental systematic since it behaves the same way in both time-interleaved data cubes. There is also a hint of a similar effect at $k_\| \sim 0.75$\,$h$\,Mpc$^{-1}$, possibly visible in Figure \ref{fig:fourier_diags} as well. We attribute both to bandpass miscalibration due to cable reflections, complicated at these frequency scales by the imperfect channelization of the MWA's two-stage polyphase filter, as well as slight antenna dependence of the bandpass due to cable length variation \cite{HazeltonEppsilon}.

\begin{figure}[] 
	\centering 
	\includegraphics[width=.48\textwidth]{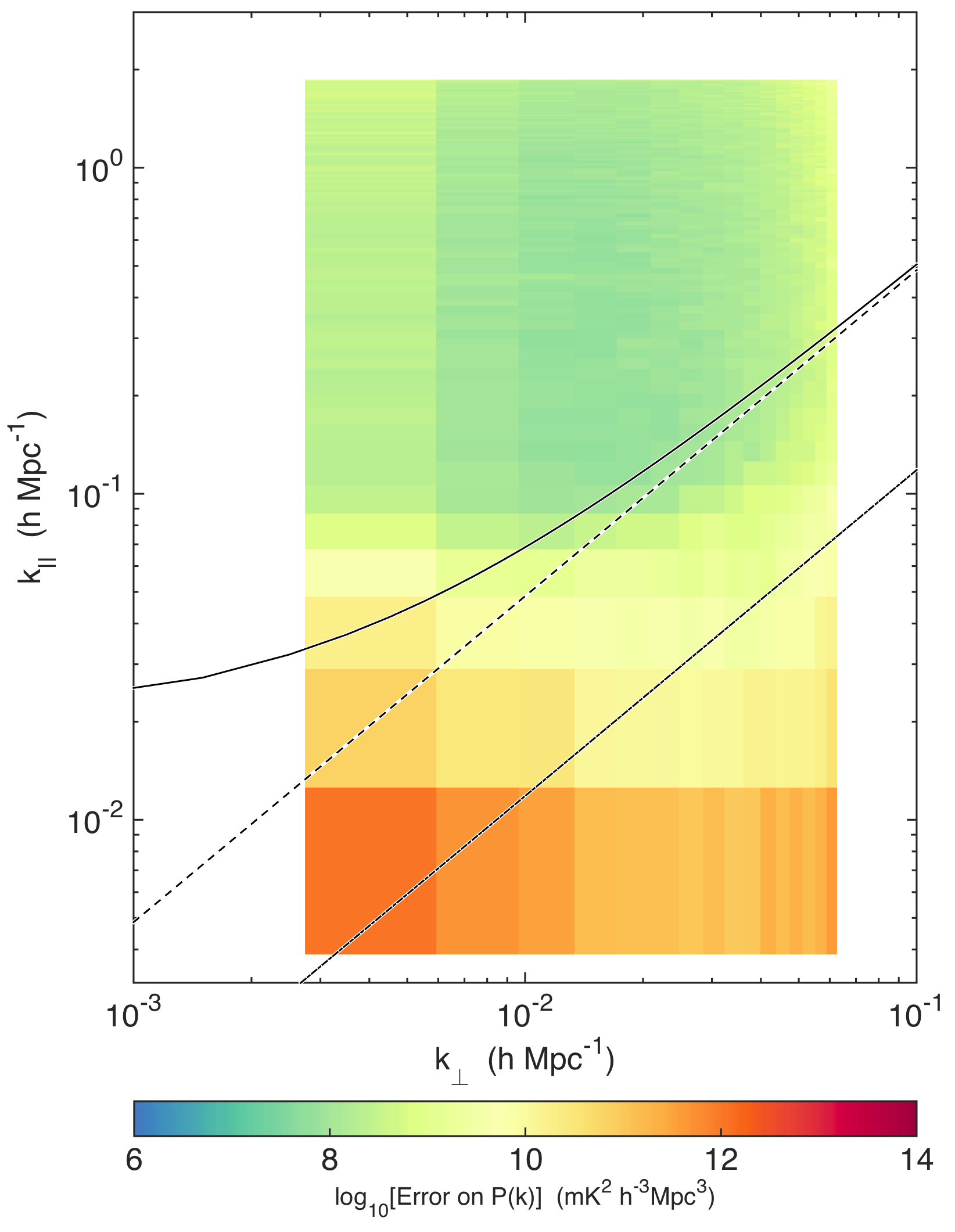}
	\caption{By including both residual foregrounds and noise in $\C$, our model for the covariance, we can calculate the expected variance on each band power in $\widehat{\p}$, which we show here. We see more variance at high (and also very low) $k_\perp$ where we have few baselines. We also see high variance at low $k_\|$ consistent with foregrounds. We see the strongest foregrounds at low $k_\perp$, which implies that the residual foregrounds have a very strong diffuse component that we have much to gain from better diffuse models to subtract. We also see that foreground-associated variance extends to higher $k_\|$ at high $k_\perp$, which is exactly the expected effect from the wedge. Both these observations are consistent with the structure of the eigenmodes we saw in Figure \ref{fig:wedge_eigenvalues}. Because we have chosen a normalization of $\widehat{\p}$ such that the $\text{Cov}(\widehat{\p})$ is diagonal, this is a complete description of our errors. Furthermore, it means that the band powers form a mutually exclusive and collectively exhaustive set of measurements.}
	\label{fig:2dError}
\end{figure}

Additionally, the quadratic estimator formalism relates our covariance models of residual foregrounds and noise to the expected variance in each band power \cite{LT11,DillonFast,X13}, which we plot in Figure \ref{fig:2dError}.
As we have chosen our power spectrum normalization $\M$ such that $\boldsymbol\Sigma \equiv \text{Cov}(\widehat{\p})$ is diagonal, it is sufficient to plot the diagonal of $\boldsymbol\Sigma^{1/2}$, the standard deviation of each band power. The EoR window is seen clearly here as well. There is high variance at low and high $k_\perp$ where the $uv$ coverage is poor, and also in the wedge due to foreground residuals. It is particularly pronounced in the bottom left corner, which is dominated by residual diffuse foregrounds.

\begin{figure}[] 
	\centering 
	\includegraphics[width=.48\textwidth]{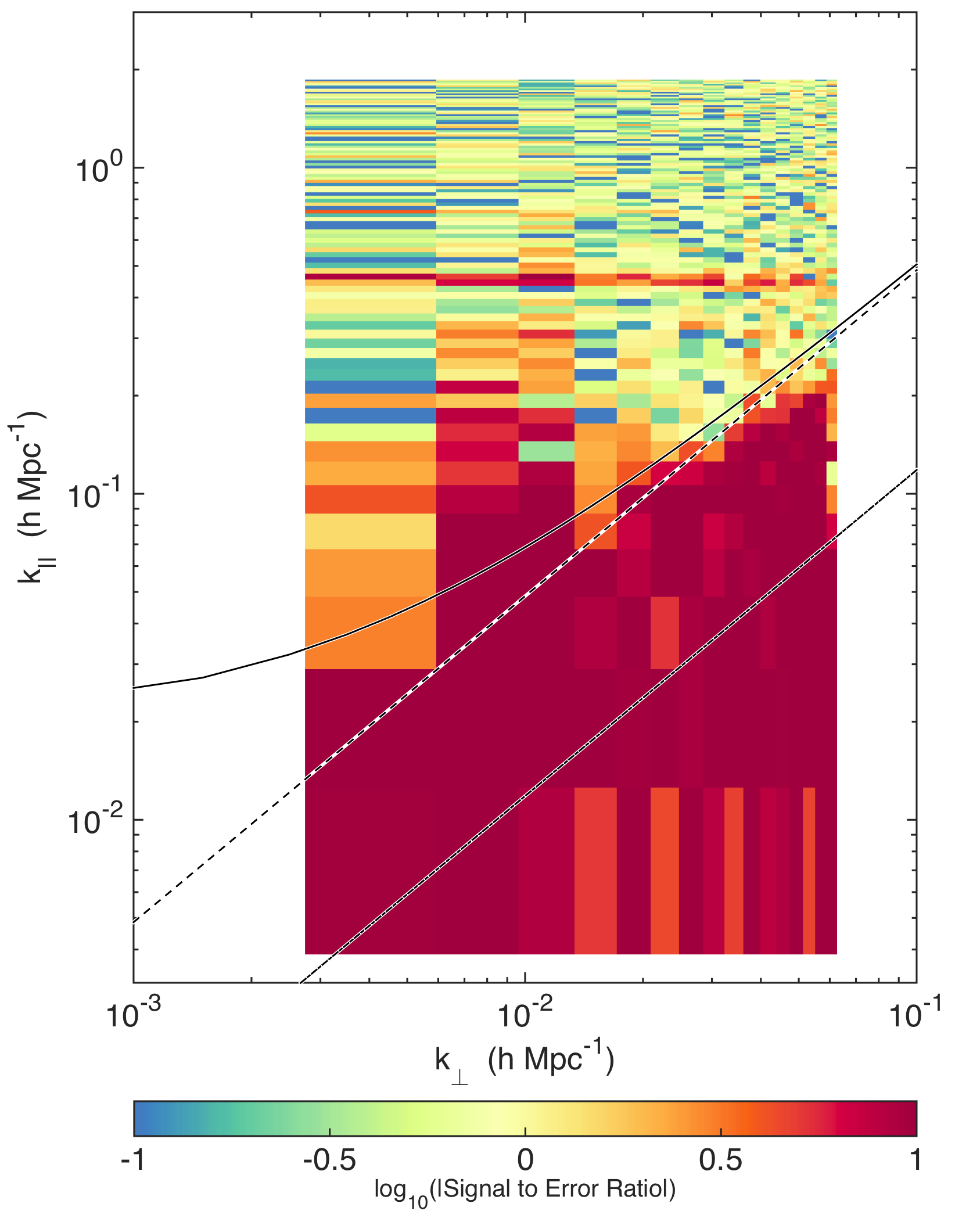}
	\caption{The foregrounds' wedge structure is particularly clear when looking at the ratio of our measured power spectrum to the modeled variance, shown here. Though the variance in foreground residual dominated parts of the $k_\perp$-$k_\|$ plane are elevated (see Figure \ref{fig:2dError}), we still expect regions with signal to error ratios greater than one. This is largely due to the fact that we choose not to subtract a foreground bias for fear of signal loss. This figure shows us most clearly where the foregrounds are important and, as with Figure \ref{fig:2dPkArcSinh}, it shows where we can hope to do better with more integration time and where we need better calibration and foreground modeling to further integrate down.}
	\label{fig:2dSNR}
\end{figure}

As our error covariance represents the error due to both noise and foregrounds we expect to make in an estimate of the 21\,cm signal, it is interesting to examine the ``signal to error ratio'' in Figure \ref{fig:2dSNR}---the ratio of Figure \ref{fig:2dPk} to Figure \ref{fig:2dError}.
The ratio is of order unity in noise dominated regions---though it is slightly lower than what we might naively expect due to our conservative estimate of $\boldsymbol\Sigma$ \cite{X13}. That explains the number of modes with very small values in Figure \ref{fig:2dError}. In the wedge and just above it, however, the missubtracted foreground bias is clear, appearing as a high significance ``detection'' of the foreground wedge in the residual foregrounds. The bandpass miscalibration line at $k_\| \sim 0.45$\,$h$\,Mpc$^{-1}$ also appears clearly due to both foreground bias and possibly an underestimation of the errors. Hedging against this concern, we simply project out this line from our estimator that bins 2D power spectra into 1D power spectra by setting the variance of those bins to infinity.

Though useful for the careful evaluation of our techniques and of the instrument, the large bandwidth data cubes used to make Figures \ref{fig:2dPk} and \ref{fig:2dPkArcSinh} encompass long periods of cosmic time over which the 21\,cm power spectrum is expected to evolve. The cutoff is usually taken to be $\Delta z \lesssim 0.5$ \cite{Yi}. These large data cubes also violate the assumption in \cite{DillonFast} that channels of equal width in frequency correspond to equal comoving distances, justifying the use of the fast Fourier transform. Therefore, we break the full bandwidth into three 10.24\,MHz segments before forming spherically averaged power spectra, and estimate the foreground residual covariance and power spectrum independently from each. We bin our 2D power spectra into 1D power spectra using the optimal estimator formalism of \cite{X13}. In our case, since we have chosen $\M$ such that $\boldsymbol\Sigma$ is diagonal, this reduces to simple inverse variance weighting with the variance on modes outside the EoR window or in the $k_\| \sim 0.45$\,$h$\,Mpc$^{-1}$ line set to infinity.

\begin{figure*}[] 
	\centering 
	\includegraphics[width=1\textwidth]{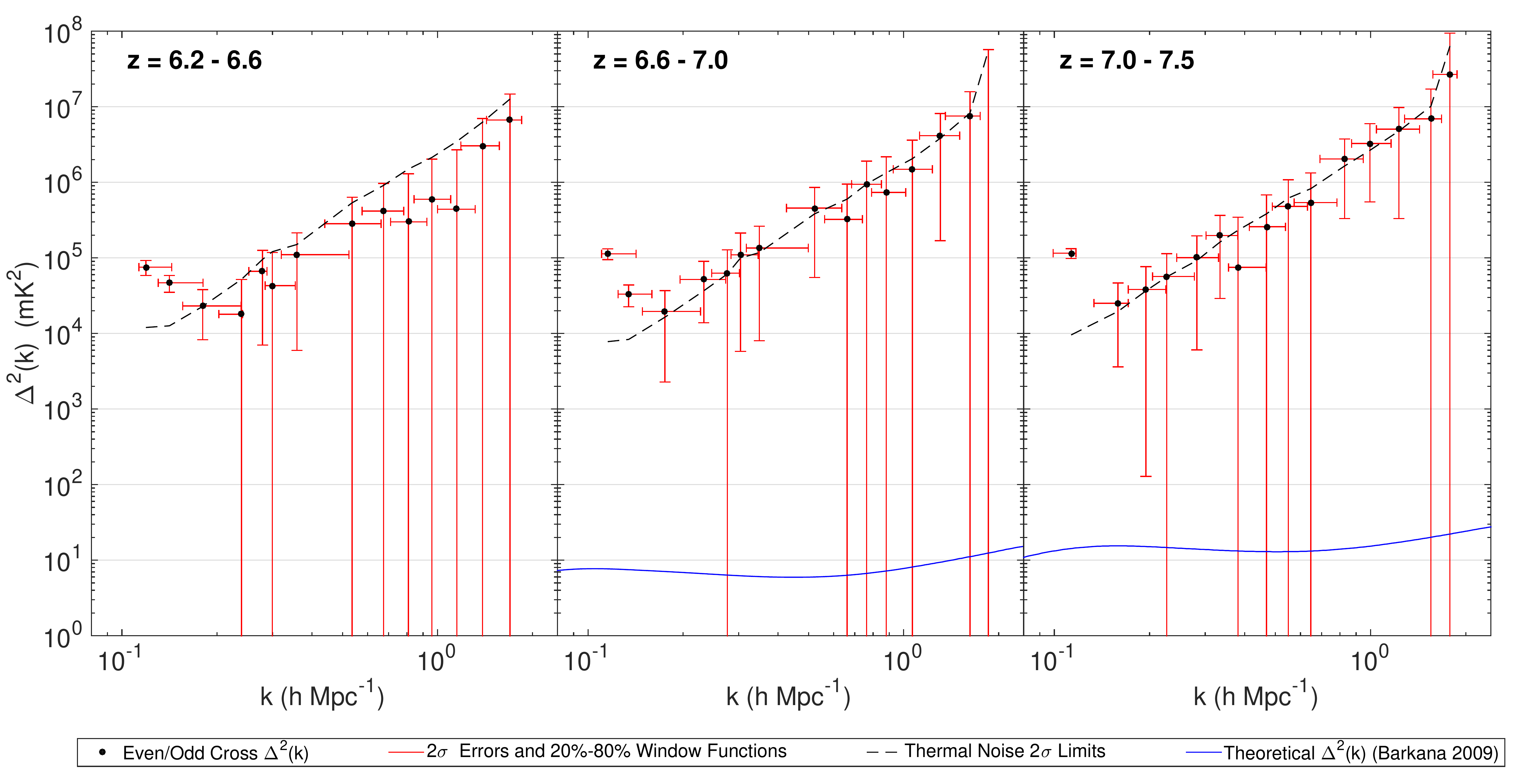}
	\caption{Finally, we can set confident limits on the 21\,cm power spectrum at three redshifts by splitting our simultaneous bandwidth into three 10.24\,MHz data cubes. The lowest $k$ bins show the strongest ``detections,'' though they are attributable to suprahorizon emission \cite{PoberWedge} that we expect to appear because we only cut out the wedge and a small buffer (0.02\,$h$\,Mpc$^{-1}$) past it. We also see marginal ``detections'' at higher $k$ which are likely due to subtle bandpass calibration effects like cable reflections. The largest such error, which occurs at bins around $k_\| \sim 0.45$\,$h$\,Mpc$^{-1}$ and can be seen most clearly in Figure \ref{fig:2dSNR}, has been flagged and removed from all three of these plots. Our absolute lowest limit requires $\Delta^2(k) < \limitDeltaSq$\,mK$^2$ at 95\% confidence at comoving scale $k = \limitk$\,$h$\,Mpc$^{-1}$ and $z = \limitz$, which is consistent with published limits \cite{newGMRT,X13,PAPER32Limits,DannyMultiRedshift,PAPER64Limits}. We also include a simplistic thermal noise calculation (dashed line), based on our observed system temperature. Though it is not directly comparable to our measurements, since it has different window functions, it does show that most of our measurements are consistent with thermal noise. For comparison, we also show the theoretical model of \cite{BarkanaPS2009} (which predicts that reionization ends before $z=6.4$) at the central redshift of each bin. While we are still orders of magnitude away from the fiducial model, recall that the noise in the power spectrum scales inversely with the integration time, not the square root.}
	\label{fig:1dDeltaSq}
\end{figure*}

In Figure \ref{fig:1dDeltaSq} we show the result of that calculation as a ``dimensionless'' power spectra $\Delta^2(k) \equiv k^3 P(k) /2\pi^2$.
We choose our binning such that the window functions (calculated as in \cite{X13} from our covariance model) were slightly overlapping. 

Our results are largely consistent with noise. Since noise is independent of $k_\|$ and $k\approx k_\|$ for most modes we measure, the noise in $\Delta^2(k)$ scales as $k^3$. We see deviations from that trend at low $k$ where modes are dominated by residual foreground emission beyond the horizon wedge and thus show elevated variance and bias in comparison to modes at higher $k$. Since we do not subtract a bias, even these ``detections'' are upper limits on the cosmological signal.

A number of barely significant ``detections'' are observed at higher $k$. Though we excise bins associated with the $k_\| \sim 0.45$\,$h$\,Mpc$^{-1}$ line, the slight detections may be due to leakage from that line. At higher $z$, the feature may due to reflections from cables of a different length, though some may be plausibly attributable to noise. Deeper integration is required to investigate further. 

Our best upper limit at 95\% confidence is $\Delta^2(k) < \limitDeltaSq$\,mK$^2$ at $k = \limitk$\,$h$\,Mpc$^{-1}$ around $z = \limitz$. Our absolute lowest limit is about 2 times lower than the best limit in \cite{X13}, though the latter was obtained at substantially higher redshift and lower $k$, making the two somewhat incomparable. Our best limit is roughly 3 orders of magnitude better than the best limit of \cite{X13} over the same redshift range, and the overall noise level (as measured by the part of the power spectrum that scales as $k^3$) is more than 2 orders of magnitude smaller. This cannot be explained by more antenna tiles alone; it is likely that the noise level was overestimated in \cite{X13} due to insufficiently rapid time interleaving of the data cubes used to infer the overall noise level.

Although one cannot directly compare limits at different values of $k$ and $z$, our limit is similar to the GMRT limit \cite{newGMRT}, $6.2\times 10^4$\,mK$^2$ at $k = 0.50$\,$h$\,Mpc$^{-1}$ and $z=8.6$ with 40\,h of observation, and remains higher than the best PAPER limit \cite{PAPER64Limits} of 502\,mK$^2$ between $k = 0.15$\,$h$\,Mpc$^{-1}$ and $k=0.50$\,$h$\,Mpc$^{-1}$ and $z=8.4$ with 4.5 months of observation.
 
In Figure \ref{fig:1dDeltaSq} we also plot a theoretical model from \cite{BarkanaPS2009} predicting that reionization has ended by the lowest redshift bin we measure. We remain more than 3 orders of magnitude (in mK$^2$) from being able to detect that particular reionization model, naively indicating that roughly 3000\,h of data are required for its detection. This appears much larger than what previous sensitivity estimates have predicted for the MWA (e.g.~\cite{MWAsensitivity}) in the case of idealized foreground subtraction. 
 
However, much of this variance is due to the residual foregrounds and systematics in the EoR window identified by our empirical covariance modeling method, not thermal noise (see Figure \ref{fig:2dError}). More integration will not improve those modes unless it allows for a better understanding of our instrument, better calibration, and better foreground models---especially of diffuse emission which might contaminate the highly sensitive bottom left corner of the EoR window. Eliminating this apparent ``suprahorizon'' emission, seen most clearly as detections in Figure \ref{fig:2dSNR} below $k \approx 0.2$\,$h$\,Mpc$^{-1}$, is essential to achieving the forecast sensitivity of the MWA \cite{MWAsensitivity}. If we can do so, we may still be able to detect the EoR with 1000\,h or fewer. This is especially true if we can improve the subtraction of foregrounds to the point where we can work within the wedge, which can vastly increase the sensitivity of the instrument \cite{MWAsensitivity,PoberNextGen}. On the other hand, more data may reveal more systematics lurking beneath the noise which could further diminish our sensitivity.


\section{Summary and Future Directions} \label{sec:summary}

In this work, we developed and demonstrated a method for empirically deriving the covariance of residual foreground contamination, $\C^{FG}$, in observations designed to measure the 21\,cm cosmological signal. Understanding the statistics of residual foregrounds allows us to use the quadratic estimator formalism to quantify the error associated with missubtracted foregrounds and their leakage into the rest of the EoR window. Because of the complicated interaction between the instrument and the foregrounds, we know that the residual foregrounds will have complicated spectral structure, especially if the instrument is not perfectly calibrated. By deriving our model for $\C^{FG}$ empirically, we could capture those effects faithfully and thus mitigate the effects of foregrounds in our measurement (subject to certain caveats which we recounted in Section \ref{sec:caveats}).

Our strategy originated from the assumption that the frequency-frequency covariance, modeled as a function of $|u|$, is the most important component of the foreground residual covariance. We therefore used sample covariances taken in annuli in Fourier space as the starting point of our covariance model. These models were adjusted to avoid double counting the noise variance and filtered in Fourier space to minimize the effect of noise in the empirically estimated covariances. Put another way, we combined our prior beliefs about the structure of the residual foregrounds with their observed statistics in order to build our models.

We demonstrated this strategy through the power spectrum analysis of a 3\,h preliminary MWA data set. We saw the expected wedge structure in both our power spectra and our variances. We saw that most of the EoR window was consistent with noise, and we understand why residual foregrounds and systematics affect the regions that they do. We were also able to set new MWA limits on the 21\,cm power spectrum from $z=6.2$ to $7.5$, with an absolute best 95\% confidence limit of $\Delta^2(k) < \limitDeltaSq$\,mK$^2$ at $k = \limitk$\,$h$\,Mpc$^{-1}$ and $z = \limitz$, consistent with published limits \cite{PAPER32Limits,DannyMultiRedshift}.

This work suggests a number of avenues for future research. Of course, improved calibration and mapmaking fidelity---especially better maps of diffuse Galactic structure---will improve power spectrum estimates and and allow deeper integrations without running up against foregrounds or systematics. Relaxing some of the mapmaking and power spectrum assumptions discussed in Section \ref{sec:caveats} may further mitigate these effects. A starting point is to integrate the mapmaking and statistical techniques of \cite{mapmaking} with the fast algorithms of \cite{DillonFast}. The present work is based on the idea that it is simpler to estimate $\C^{FG}$ from the data than from models of the instrument and the foregrounds. But if we can eliminate systematics to the point where we really understand $\PSF$, the relationship between the true sky and our dirty maps, then perhaps we can refocus our residual foreground covariance modeling effort on the statistics of the true sky residuals using the fact that $\C^{FG} = \PSF \C^{FG,\text{true}} \PSF^\trans.$ Obtaining such a complete understanding of the instrument will be challenging, but it may be the most rigorous way to quantify the errors introduced by missubtracted foregrounds and thus to confidently detect the 21\,cm power spectrum from the epoch of reionization.

\section*{Acknowledgments}
We would like to thank Adrian Liu, Aaron Parsons, and Jeff Zheng for helpful discussions. We also acknowledge an anonymous referee whose insightful questions led to the refinement of method in order to a bias that occurs when a $uv$ cell is used to estimate its own covariance and the ultimate form of Equation \ref{eq:CovEstOtherAnnuli}. We would also like to thank Rennan Barkana for the theoretical power spectra in Figure \ref{fig:1dDeltaSq}. This work was supported by NSF Grants AST-0457585, AST-0821321, AST-1105835, AST-1410719, AST-1410484, AST-1411622, and AST-1440343, by the MIT School of Science, by the Marble Astrophysics Fund, and by generous donations from Jonathan Rothberg and an anonymous donor. D.C.J. would like to acknowledge NSF support under AST-1401708.

This scientific work makes use of the Murchison Radio-astronomy Observatory, operated by CSIRO. We acknowledge the Wajarri Yamatji people as the traditional owners of the Observatory site. Support for the MWA comes from the U.S. National Science Foundation (grants AST-0457585, PHY-0835713, CAREER-0847753, and AST-0908884), the Australian Research Council (LIEF grants LE0775621 and LE0882938), the U.S. Air Force Office of Scientific Research (grant FA9550-0510247), and the Centre for All-sky Astrophysics (an Australian Research Council Centre of Excellence funded by grant CE110001020). Support is also provided by the Smithsonian Astrophysical Observatory, the MIT School of Science, the Raman Research Institute, the Australian National University, and the Victoria University of Wellington (via grant MED-E1799 from the New Zealand Ministry of Economic Development and an IBM Shared University Research Grant). The Australian Federal government provides additional support via the Commonwealth Scientific and Industrial Research Organisation (CSIRO), National Collaborative Research Infrastructure Strategy, Education Investment Fund, and the Australia India Strategic Research Fund, and Astronomy Australia Limited, under contract to Curtin University. We acknowledge the iVEC Petabyte Data Store, the Initiative in Innovative Computing and the CUDA Center for Excellence sponsored by NVIDIA at Harvard University, and the International Centre for Radio Astronomy Research (ICRAR), a Joint Venture of Curtin University and The University of Western Australia, funded by the Western Australian State government.

\bibliography{Empirical_Covariance}  

\end{document}